\documentstyle[art12]{article}
\input{epsf}
\newcommand{\be}{\begin{equation}}
\newcommand{\ee}{\end{equation}}
\newcommand{\ba}{\begin{eqnarray}}
\newcommand{\ea}{\end{eqnarray}}
\newcommand{\bea}{\begin{eqnarray*}}
\newcommand{\bet}{\begin{center} \begin{tabular}}
\newcommand{\eea}{\end{eqnarray*}}
\newcommand{\ent}{\end{tabular} \end{center}}
\newcommand{\bb}{\begin{thebibliography}}
\newcommand{\eb}{\end{thebibliography}}
\newcommand{\ci}[1]{\cite{#1}}
\newcommand{\bi}[1]{\bibitem{#1}}

\newcommand{\ra}{\rightarrow}
\newcommand{\re}[1]{(\ref{#1})}

\newcommand{\bit}{\begin{itemize}}
\newcommand{\eit}{\end{itemize}}
\newcommand{\al}{\alpha}
\newcommand{\g}{\gamma}

\newcommand{\Gmu}{G_\mu}

\newcommand{\crn}{\nonumber \\}
\newcommand{\nn}{\nonumber}

\newcommand{\mz }{M_Z^2}

\newcommand{\dal}{\Delta \alpha}
\newcommand{\dalh}{\Delta \alpha^{(5)}_{\rm had}}

\newcommand{\sinW}{\sin^2 \Theta_W}
\newcommand{\siqW}{\sin^4 \Theta_W}

\newcommand{\wz}{\sqrt{2}}
\newcommand{\SM}{Standard Model }

\newcommand{\epm}{e^+e^-}
\newcommand{\epml}{e^+e^- \rightarrow \ell^+\ell^-}
\newcommand{\veps}{\varepsilon}
\newcommand{\half}{\frac{1}{2}}
\newcommand{\npb}[1]{{\em Nucl.\ Phys.\ }{\bf B{#1}} }
\newcommand{\np}[1]{{\em Nucl.\ Phys.\ }{\bf B{#1}} }
\newcommand{\plb}[1]{{\em Phys.\ Lett.\ }{\bf B{#1}} }       
\newcommand{\pl}[1]{{\em Phys.\ Lett.\ }{\bf {#1}B} }        
\newcommand{\prd}[1]{{\em Phys.\ Rev.\ }{\bf D{#1}} }
\newcommand{\PR}[1]{{\em Phys.\ Rep.\ }{\bf C{#1}} }
\newcommand{\prl}[1]{{\em Phys.\ Rev.\ Lett.\ }{\bf {#1}} }
\newcommand{\pr}[1]{{\em Phys.\ Rev.\ }{\bf {#1}} }
\newcommand{\zp}[1]{{\em Z.\ Phys.\ }{\bf C{#1}} }
\newcommand{\JL}[1]{{\em JETP \ Lett. \ }{\bf {#1}} }
\newcommand{\lnc}[1]{{\em Lett. Nuovo Cim. \ }{\bf {#1}} }

\newcommand{\SJNP}[1]{{\em Sov. J. Nucl. Phys. \ }{\bf {#1}} }
\newcommand{\IJMP}[1]{{\em Int. J. Mod. Phys. A \ }{\bf {#1}} }
\begin{document}

\newcommand{\mysymb}[5]{
  \unitlength1mm \begin{picture}(#4,0) \put(0,0){ 
 }
\end{picture} }
\parindent 0mm
\parskip 2mm
\renewcommand{\arraystretch}{1.4}
\thispagestyle{empty}
\hfill {\sc PSI-PR-95-1} \qquad { } \par
\hfill {\sc BudkerINP 95-5} \qquad { } \par
\hfill January 1995 \qquad { }

\vspace*{15mm}

\begin{center}
\renewcommand{\thefootnote}{\fnsymbol{footnote}}
{\LARGE Hadronic contributions to $(g-2)$ of the leptons and to the
effective fine structure constant $\alpha(\mz)$}

\vspace{16mm}

{\large S.~Eidelman}

\medskip

{\em Budker Institute for Nuclear Physics, 630090 Novosibirsk, Russia}
\bigskip

{\large F.~Jegerlehner}

\medskip

{\em Paul Scherrer Institute, CH-5232 Villigen PSI, Switzerland}
\bigskip

\vspace*{20mm}

\textwidth 120mm
\begin{abstract}

The new experiment planned at Brookhaven to measure the anomalous
magnetic moment of the muon $a_\mu \equiv (g_\mu-2)/2$ will improve
the present accuracy of 7 ppm by about a factor of 20. This requires a
careful reconsideration of the theoretical uncertainties of the $g-2$
predictions, which are dominated by the error of the contribution from
the light quarks to the photon vacuum polarization.  This issue is
crucial also for the precise determination of the running fine
structure constant at the $Z$--peak as LEP/SLC experiments continue to
increase their precision.  In this paper we present an updated
analysis of the hadronic vacuum polarization using all presently
available $\epm$ data.  This seems to be justified because previous
work on the subject was based to some extent on preliminary or
incomplete experimental data.  Contributions from different energy
ranges are presented separately for $g-2$ of the muon and the
$\tau$--lepton and for $\alpha(\mz)$. We obtain the results
$a_\mu^{{\rm had}\:*}=(725
\pm 16)\times 10^{-10}$ and $a_\tau^{{\rm had}\:*}=(351 \pm 10)\times
10^{-8}$, where the asterisk indicates the dressed (renormalization
group improved) value. For the effective fine structure constant at
$M_Z=91.1888$ GeV we obtain $\dalh = 0.0280 \pm 0.0007 $ and $ \al
(\mz)^{-1}=128.896 \pm 0.090$.  Further improvement in the accuracy of
theoretical predictions which depend on the hadronic vacuum
polarization requires more precise measurements of $\epm$
cross--sections at energies below about 12 GeV in future experiments.

\end{abstract}
\end{center}
\bigskip

\textwidth 170mm
\newpage

\setcounter{page}1

\renewcommand{\thefootnote}{\arabic{footnote}}
\setcounter{footnote}{0}

\section{Introduction}

The anomalous magnetic moment of the muon has been measured with very
high precision at the CERN Muon Storage Ring~\ci{amuexp}.  It is one
of the best measured quantities in physics. As it can be calculated
with high accuracy \ci{Kino1,KiLi,BKT} it provides an extremely clean
test of electroweak theory and may give us important hints on possible
deviations from the \SM (SM)\ci{KiMa,Mars,Bernr,AEW}. The special
interest comes about because $a_\ell \;(\ell=e, \mu, \tau)$
corresponds to a helicity flip coupling $\bar{\ell}_L \sigma_{\mu \nu}
F^{\mu \nu}\ell_R$ which must vanish at the tree level for any fermion
in any renormalizable field theory. In the SM it is thus a finite
calculable quantity nonvanishing only due to quantum fluctuations.
Another interesting feature is that $a_\ell$ is finite only if the
$WW\gamma$--coupling is of the Yang-Mills type. In the limit $m_\ell
\ra 0$ regular contributions to $a_\ell$ vanish in the SM because $\ell_L$
does not couple to $\ell_R$ in this limit. This brings one power of
$m_\ell$ into the effective $\bar{\ell} \ell \gamma$ vertex; a second
power is due to the normalization to the lepton Bohr magneton
$e/2m_\ell$.

The $m_\ell^2$--dependence of the weak interaction and the vacuum
polarization effects makes them completely unobservable for the
electron. The effects are enhanced however by the large factor
$m^2_\mu / m^2_e$ for the muon relative to the electron.  This
enhancement is very welcome and happens for all kinds of new physics
that could couple to photons and leptons. Therefore $a_\mu$ is an
important tool for obtaining stringent upper bounds on new physics
contributions. For the $\tau$ (see e.g. \ci{tau1,tau2}) there is an
additional enhancement factor $m^2_\tau / m^2_\mu$ which magnifies the
interesting effects. However, because of the short $\tau$--lifetime, a
measurement of $a_\tau$ is very difficult. Therefore one is still off
by a few orders of magnitude in establishing its value experimentally.
The best possibility to determine $a_\tau$ is by the leptonic
radiative $\tau$--decay $\tau^- \ra e^- \bar{\nu}_e
\nu_\tau \gamma$ which also provides the best present upper bound \ci{L3}.

The present accuracy \ci{amuexp} $$a_\mu^{{\rm exp}}=(11659 230 \pm 85
)\times 10^{-10}$$ allows us to test QED with very high precision. The
theoretical prediction includes a number of terms of different origin.
We may write

$$a_\mu^{{\rm the}} =a_\mu^{{\rm QED}}+a_\mu^{{\rm had}}+a_\mu^ {\rm
weak} +a_\mu^{{\rm new}}\;,$$

where the first, and by far largest term, is from pure QED,
$a_\mu^{{\rm had}}$ denotes the virtual hadronic (quark) contribution,
which will be studied in this paper, $a_\mu^{{\rm weak}}$ summarizes
the SM effects due to virtual $W$, $Z$ and Higgs particle exchanges
and $a_\mu^{{\rm new}}$ stands for possible contributions from
extensions of the SM. If we assume the last term to be zero we have
the theoretical prediction \ci{Kino1}

$$a_\mu^{{\rm the}} = (11 659 192 \pm 18)\times 10^{-10}\;,$$

and therefore at present we have $$a_\mu^{{\rm exp}}-a_\mu^{{\rm
the}}=(38 \pm 87)\times 10^{-10}\;.$$

With the precision attempted by the forthcoming Brookhaven
experiment~\ci{BNL} one should be able to establish the weak SM
contribution $a_\mu^{{\rm weak}}\simeq$20$\times$10$^{-10}$, for
example.  In fact, however, such contributions may be concealed by the
theoretical uncertainty which is of a similar size. The latter is
dominated by the uncertainty of the hadronic contribution and
therefore a careful analysis of the problem is of primary importance.
The problem is that the low energy hadronic effects cannot be
calculated in perturbative QCD and one has to rely on the
semi-phenomenological dispersion theoretical approach
\ci{BM,Durand,GdeR} which allows us to compute $a_\mu^{{\rm had}}$ as
an integral over experimental data from $\epm$ annihilation. The
experimental errors of the data of course imply then a theoretical
uncertainty of $g-2$ predictions.

A comparison of different results based at least partially on this
approach is given in Tab.~1 which illustrates the present status.
\bet{|l|lc|}
\multicolumn{3}{c}{Table~1: Comparison of estimates of
$a_\mu^{\rm had}\cdot 10^{10}$ by different authors}\\
\hline
 $a_{\mu}^{\rm had}\cdot 10^{10}$ & Author & Year [Reference] \\
\hline
663 $\pm$ 85 & Barger et al. & 1975 \ci{Barger75} \\ 684 $\pm$ 11 &
Barkov et al. & 1985 \ci{Barkov85} \\ 707 $\pm$ ~6 $\pm$ 16 &
Kinoshita et al. & 1985 \ci{Kino2} \\ 710 $\pm$ 10 $\pm$ ~5 & Casas et
al. & 1985 \ci{CLY} \\ 705 $\pm$ ~6 $\pm$ ~5 & Martinovi\v{c},
Dubni\v{c}ka & 1990 \ci{MD} \\ 724 $\pm$ ~7 $\pm$ 26 & Jegerlehner &
1991 \ci{jegup} \\ 699 $\pm$ ~4 $\pm$ ~2 & Dubni\v{c}kov\'{a} et al. &
1992 \ci{DD} \\ 702 $\pm$ ~6 $\pm$ 14 & \multicolumn{1}{c}{This \
work} &\\ 725 $\pm$ ~6 $\pm$ 15 & \multicolumn{1}{c}{RG improved} &\\
\hline
\ent

The second quantity considered in this paper is the hadronic
contribution $\dalh (\mz)$ of the 5 light quark flavors to the shift
of the fine structure constant from the Thomson limit to the
$Z$--resonance $\dal=1-\alpha/\alpha(\mz)$. This quantity can also
only be estimated reliably in terms of the experimental $\epm$ data
with corresponding uncertainties. The effective fine structure
constant $\alpha(s)$ plays a crucial role in precision physics at all
energy scales beyond the very low energy region. It is particularly
important for all physics of the weak gauge bosons, as being studied
currently at LEP and SLC.  Almost all SM predictions of observables in
terms of $\al$, $\Gmu$ and $M_Z$, the most precise set of input
parameters available, depend on $\dal(s)$.  Again, the uncertainty of
$\dalh (\mz)$ is a limiting factor which obscures the interpretation
of precision measurements at a certain level. In fact, for the
leptonic effective weak mixing parameter $\sin^2 \theta_{{\rm
eff}}^{\rm lept} (M_Z)$, which is determined in $\epml$ at the
$Z$-resonance, the LEP collaboration \ci{LEP} has reached an
experimental error $\delta
\sin^2 \theta_{{\rm eff}}^{\rm lept} \simeq 0.0004$ which is only
slightly larger than the hadronic uncertainty of about 0.0003 of the
SM prediction for this quantity. It thus has become a very crucial
question whether it is possible to reduce the error of $\dalh (\mz)$
further and what the perspectives are in the future. The present
situation is summarized by the results in Tab.~2.

Note that some of the values have been shifted by
$$\frac{\al}{3\pi}\frac{22}{3}\:(1+\frac{\al_s}{\pi})\:\ln
(M'_Z/M_Z)$$ from the pre--LEP reference value $M_Z=93$ GeV to the
current value of $M_Z$.  Ref.~\ci{LynnPensoVerz} quotes $\dalh
(-Q_0^2)=0.0145 \pm 0.0012$ for $Q_0^2=79$ GeV$^2$. We have added
$\dalh (M_Z^2)-\dalh (-Q_0^2)=0.0138$ which is obtained by using 3rd
order perturbative QCD.  For a comparison of the earlier
results~\ci{Berends76} we refer to \ci{Jegerlehner86}.  The
differences are mainly due to a different treatment of the systematic
errors and/or different model assumptions.  The present situation
justifies a reconsideration of the problem.

\bet{|l|lc|}
\multicolumn{3}{c}{Table~2: Comparison of estimates of
$\dalh (\mz)$ by different authors}\\
\hline
$\dalh (\mz)$ & Author & Year [Reference] \\
\hline
0.0285~ $\pm$ 0.0007 & Jegerlehner & 1986 \ci{Jegerlehner86} \\
0.0283~ $\pm$ 0.0012 & Lynn et al. & 1987 \ci{LynnPensoVerz} \\
0.0287~ $\pm$ 0.0009 & Burkhardt et al. & 1989 \ci{Burkhardt89} \\
0.0282~ $\pm$ 0.0009 & Jegerlehner & 1991 \ci{TASI} \\ 0.02666 $\pm$
0.00075& Swartz & 1994 \ci{Swartz94} \\ 0.02732 $\pm$ 0.00042& Martin
and Zeppenfeld & 1994 \ci{MZ94} \\ 0.0280~ $\pm$ 0.0007 &
\multicolumn{1}{c}{This \ work} &\\
\hline
\ent

    To calculate the necessary dispersion integrals below we prefer to
use direct integration over the experimental values of cross sections.
In this approach one can take into account uncertainties of separate
measurements in a straightforward manner. The alternative method which
was used in most of the previous works is to make a fit of the
experimental points within some model and integrate the arising
parametrization of the data. This procedure inevitably leads to a
model dependence and it is not clear how experimental errors
especially systematic uncertainties can be taken into account.

The two recent papers~\ci{Swartz94,MZ94} appeared while we were
writing up this update. We have added some remarks which should
clarify at least part of the discrepancies between these papers and
the present analysis which yields results consistent with previous
ones (e.g.~\ci{TASI}).

In Secs.~2 and 3 we will briefly review how the hadronic contributions
to $a_\ell$ and $\dal$ are determined by the $\epm$ data.  Details on
the evaluation of the dispersion integrals and the treatment of the
errors are discussed in Sec.~4. In Sec.~5 we describe the $\epm$ data
which we use in our analysis and in Sec.~6 we present the results. A
brief summary and outlook follows in Sec.~7.

\section{Hadronic contributions to the anomalous magnetic moments of the
leptons}

The leading hadronic contribution to $g-2$ is due to the photon vacuum
polarization insertion into the vertex diagram of the electromagnetic
vertex of a lepton. The corresponding diagram is shown in Fig.~1.

\vspace*{20mm}

\bea
\mysymb{0}{0}{40}{40}{gm2had}
\eea

\vspace*{-38mm}

\begin{picture}(60,60)(-52,0)
\put(20,10){$\gamma$}
\put(40,10){$\gamma$}
\put(28, 5){had}
\put(33,44){$\gamma$}
\put( 9, 7){$\ell$}
\put(52, 7){$\ell$}
\end{picture}

\vspace*{-5mm}

\begin{center}
\begin{minipage}[h]{12.3cm} \baselineskip 10truept
\begin{tabular}{ll}
{\bf Fig.~1:} &
\begin{tabular}[t]{p{100mm}}
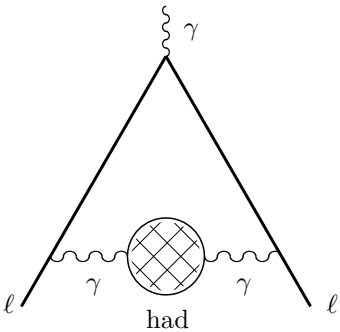
Leading hadronic vacuum polarization contribution to $g-2$ of a
lepton.
\end{tabular} \end{tabular}
\end{minipage}
\end{center}

\vspace*{5mm}

The ``blob'' represents the irreducible photon self-energy.
Subleading hadronic contributions are obtained by multiple self-energy
insertions or insertions of irreducible light-by-light scattering
(4$\gamma$) or higher amplitudes \ci{Kino2,Barbi}. We will not discuss
such subleading terms in this paper unless stated otherwise.

As already mentioned before, the low energy hadron effects which are
needed here, and in principle are determined by QCD, cannot be
obtained by using perturbation theory.  Fortunately, this contribution
may be calculated in terms of the experimental total cross section
$\sigma_{had}=\sigma(\epm \ra {\rm hadrons})$ of $\epm$ annihilation
into any hadronic state by using the familiar dispersion integral
\ci{BM,Durand,GdeR}
\be
a_\mu^{had} =\frac{1}{4\pi^3}
\int_{4m^2_\pi}^{\infty} ds\: \sigma^{(0)}_{had}(s)\:K(s)
=\left( \frac{\alpha m_\mu}{3\pi} \right)^2
\int_{4m^2_\pi}^{\infty} ds\: \frac{R(s)\:\hat{K}(s)}{s^2}
\label{amu}
\ee
which can be evaluated by using the experimental data for
$\sigma_{had}(s)$ or $R(s)$ up to some sufficiently high energy, e.g.,
$E_{\rm cut}=40$ GeV, and by perturbative QCD for the high energy
tail.

For the moment we are interested in calculating the contribution from
the irreducible photon self-energy, in which case we have to use the
``undressed'' ($\equiv$ lowest order with respect to QED) hadronic
cross section
\be
\sigma^{(0)}_{had}(s)=\sigma_{had}(s)\: (\al / \al(s))^2\;\;.
\ee
We refer to the Appendix for a brief discussion of ``dressed'' versus
``undressed'' quantities in dispersion relations.

The second form in Eq.~\re{amu} is convenient for the evaluation of
contributions at higher energies. It is an expression in terms of the
cross-section ratio
\be
R(s)=\frac{\sigma_{tot} (\epm \ra \gamma^* \ra hadrons)} {\sigma (\epm
\ra \gamma^* \ra \mu^+ \mu^-)}\;.
\label{Rdef}
\ee
Note that $R(s)$, by the proper definition according to Eq.~\re{Rdef},
is the {\em ratio of the total cross sections} which is determined by
QCD. In perturbative QCD we have \ci{Rpert3}
\be
R(s)=3\sum_f Q_f^2 \sqrt{1-4m_f^2/s}\: (1+2m_f^2/s)
\left(1+a+c_1 a^2+c_2 a^3 +\cdots \right)
\label{Rpert}
\ee
where $Q_f$ and $m_f$ denote the charge and mass of the quark,
respectively, $a=\alpha_s (s)/\pi$ with $\alpha_s (s)$ the strong
coupling constant, and
\bea
c_1&=&~~1.9857-0.1153 N_f \crn c_2&=&-6.6368-1.2002 N_f - 0.0052 N_f^2
-1.2395\: (\sum Q_f)^2/(3 \sum Q_f^2)
\eea
in the $\overline{MS}$ scheme. $N_f$ is the number of active flavors.
This result is applicable about 1 GeV above the resonances and at
sufficiently high energies and will be used in particular to calculate
the high energy tail of Eq.~\re{amu}. We will assume a top mass of
173~GeV~\ci{LEP} for the calculation of the perturbative tail. As a
check we also will compare the experimental data in the region above
the $\psi$--resonances with the calculation of $R(s)$ presented
recently in Ref.~\ci{ChetyrkinKuehn95}, which is improved by charm and
bottom mass effects.

      Usually experiments do not determine $R$ as a ratio of the total
cross sections as given by Eq.~\re{Rdef}. Rather the hadronic
experimental cross section is first corrected for QED
effects~\ci{BonneauMartin,Tsai,BerendsKleiss,Eidelman78}, which
include bremsstrahlung as well as vacuum polarization corrections. The
latter account for the running of the fine structure constant
$\alpha(s)$.  After these corrections have been applied $\sigma_{tot}$
is divided by the Born cross section $\sigma_0 (\epm \ra \gamma^* \ra
\mu^+\mu^-)=\frac{4\pi\al ^2}{3s}$ so that
\bea
R(s)=\frac{\sigma_{tot} (\epm \ra \gamma^* \ra hadrons)_{exp}^{corr}}
{\sigma_0 (\epm \ra \gamma^* \ra \mu^+ \mu^-)}\;.
\eea
Note that, the experimental cross section $\sigma (\epm \ra \gamma^*
\ra \mu^+ \mu^-)$ never appears here and is used by careful groups to 
check how good normalization is (see e.g., \ci{BerBoe}).  The question
of how $R$ has been determined precisely in a given experiment is not
always clear. We will comment on this point when discussing the data
below.

A renormalization group (RG) improvement of the result may be obtained
by resumming the multiple irreducible self-energy insertions, which is
equivalent to using the full photon propagator in Fig.~1.  As shown in
the Appendix, this is simply achieved by using the physical cross
section $\sigma_{had}(s)$ or, equivalently, the ``dressed''
$R$--function $R(s)^{dressed}=R(s) (\al (s)/ \al )^2$ under the
dispersion integral Eq.~\re{amu}.

Turning back to Eq.~\re{amu}, the kernel $K(s)$ may conveniently be
written in terms of the variable
\bea
x=\frac{1-\beta_\mu}{1+\beta_\mu}\;,\;\;\beta_\mu=\sqrt{1-4m^2_\mu/s}
\eea
and is given by
\be
K(s)=\frac{x^2}{2}\:(2-x^2)+\frac{(1+x^2)(1+x)^2}{x^2}
\left(\ln(1+x)-x+\frac{x^2}{2} \right) +\frac{(1+x)}{(1-x)}\:x^2 \ln(x)\;\;.
\label{KS}
\ee
Note that the function
\bea
\hat{K}(s)=\frac{3 s}{m^2_\mu}K(s)
\eea
is bounded: it increases monotonically from 0.63 at threshold
$s=4m^2_\pi$ to 1 at $s \ra \infty$. It should be noted that for small
$x$ the calculation of the function $K(s)$, in the form given above,
is numerically instable and we instead use the asymptotic expansion
(used typically for $x \leq 0.0006$)
\bea
K(s)=\left(\frac13+\left(\frac{17}{12}+\left(\frac{11}{30}
+\left(-\frac{1}{10}+\frac{3}{70}
x\right)\:x\right)\:x\right)\:x\right)\:x+\frac{1+x}{1-x}\:x^2\ln(x)\;.
\eea
Other representations of $K(s)$, like the simpler--looking form
\bea
K(s)=\half-r+\half r\:(r-2)\:\ln(r)+\left(1-2r+\half r^2\right)
\:\ln(x)/\beta_\mu\;\;,
\eea
with $r=s/m^2_\mu$, are much less suitable for numerical evaluation
because of much more severe numerical cancellation (even less stable
is the representation utilized in \ci{DD}).

The representation Eq.~\re{KS} of $K(s)$ is valid for the muon (or
electron) where we have $s>4m^2_\mu$ in the domain of integration
$s>4m^2_\pi$, and $x$ is real, and $0 \leq x \leq 1$. For the $\tau$
Eq.~\re{KS} applies for $s>4m^2_\tau$. In the region $4m^2_\pi < s < 4
m^2_\tau$, where $0<r=s/m^2_\tau<4$, we may use the form
\be
K(s)=\half-r+\half r\:(r-2)\:\ln(r)-\left(1-2r+\half
r^2\right)\:\varphi/w
\ee
with $ w=\sqrt{4/r-1}$ and $\varphi=2 \tan^{-1}(w)$.

\section{Hadronic contributions to the running of the effective fine structure
constant}
\label{sec:dal}

The effective fine structure constant at scale $\sqrt{s}$ is given by
\be
\alpha(s)=\frac{\alpha}{1-\dal(s)}\;\;,
\label{runalp}
\ee
where $\alpha$ is the fine structure constant and $\dal$ is the photon
vacuum polarization contribution. In terms of the one particle
irreducible photon self-energy $\Pi_{\gamma}(s)=s\Pi'_{\gamma}(s)$ we
have
\be
\dal(s)= \Pi'_{\gamma}(0)-{\rm Re}\:\Pi'_{\gamma}(s)\;\;,
\label{irrse}
\ee
which for $s=\mz$, for example, is large due to the large change in
scale going from zero momentum (Thomson limit) to the Z-mass scale
$\mu=M_Z$. In perturbation theory, the leading light fermion ($m_f \ll
M_W, \sqrt{s}$) contribution is given by

\vspace*{-5mm}

\ba
\Delta \alpha (s)&=&  \sum_f \; \mysymb{0}{0}{20}{20}{prsym2} \crn
&=& \frac{\alpha}{3\pi}\sum_f Q_{f}^2 N_{cf}(\ln \frac{s}
{m_f^2}-\frac{5}{3}\:)\;\;,
\label{dallep}
\ea

\vspace*{-34mm}

\begin{center}
\begin{picture}(80,35)(-5,0)
\put(28,31){$\gamma$}
\put(34,33){$f$}
\put(41,22){$f$}
\put(45,31){$\gamma$}
\end{picture}
\end{center}

\vspace*{-14mm}

with $Q_f$ the fermion charge and $N_{cf}$ the color factor, 1 for
leptons and 3 for quarks. We distinguish the contributions from the
leptons, for which Eq.~\re{dallep} is appropriate, the five light
quarks and the top
\be
\Delta \alpha =\Delta \al _{l}+
                 \dalh+ \Delta \al _{top}\;.
\ee
Since the top quark is heavy we cannot use the light fermion
approximation for it. A very heavy top in fact decouples like
\bea
\Delta \al _{top}\simeq -\frac{\al}{3\pi}\frac{4}{15} \frac{\mz}{m_t^2} \ra 0
\nn
\eea
when $m_t \gg M_Z$.

A serious problem is the low energy contributions of the five light
quarks u, d, s, c and b which cannot be reliably calculated by using
perturbative QCD. Fortunately, again one can evaluate this hadronic
term $\dalh$ from hadronic $\epm $--annihilation data by using a
dispersion relation together with the optical theorem which results in
the integral~\ci{CabbiboGatto61,Berends76}
\be
\dalh (M_Z^2) =
-\frac{M_Z^2}{4\pi^2 \al}{\rm \ Re}\int_{4m_{\pi}^2}^{\infty}
ds\frac{\sigma^{(0)}_{had}(s)}{s-M_Z^2-i\veps} = -\frac{\al
M_Z^2}{3\pi}{\rm
\ Re}\int_{4m_{\pi}^2}^{\infty} ds\frac{R(s)}{s(s-M_Z^2-i\veps)}\;\;,
\label{dal}
\ee
which is very similar to the one we encountered for $g-2$ in
Eq.~\re{amu}. The only difference is the different weight-function
multiplying $R(s)$ under the integral. Since Eq.~\re{amu} has an extra
factor $1/s$ at low $s$ the low energy data play a dominant role for
$a_\mu$ while Eq.~\re{dal} gets significant contributions from a broad
energy range up to about $E\simeq M_Z/2$, as we shall see below.

Note that the remarks made earlier about the proper definition of
$R(s)$ apply here as well.

Above the $\Upsilon$ energy we apply a correction factor for
$\gamma-Z$ mixing to $R(s)^{exp}$. The correction for the
$Z$--exchange contribution is given by the ratio $c_{QQ}/c_{\gamma Z}$
where
\bea
c_{QQ} = \sum_f Q_f^2=11/9\;\;;\;\;\; c_{\gamma Z} = c_{QQ} - 2 v_e
c_{Qv} P(s) + c_{e} c_{f} P(s)^2\;\;.
\eea
We denoted the $Z$--pole factor by
\bea
P(s)=\frac{\wz \Gmu M_Z^2}{16 \pi \alpha} \frac{s}{s-M_Z^2}
\eea
and the coefficients are determined by the $Zf\bar{f}$ couplings $v_f$
and $a_f$ as follows:
\bea \begin{array}{lcccl}
c_{Qv} &=& \sum_f Q_f v_f &=& 7/3-44/9 \sinW \\ c_e &=& v_e^2+a_e^2
&=& 2-8 \sinW + 16 \siqW \\ c_f &=& \sum_f (v_f^2+a_f^2)&=&10-56/3
\sinW +176/9 \siqW \;\;.
\end{array}
\eea
In our normalization $v_e=-1+4 \sinW$. We use the LEP values
$M_Z=91.1888$ GeV and $\sinW=\sin^2 \theta_{{\rm eff}}^{\rm lept}
(M_Z)=0.2322$~\ci{LEP} for numerical estimates.

\section{Evaluation of the dispersion integral}

In order to obtain a conservative estimate of the relevant integrals
we try to rely on the experimental data as much as possible and
integrate directly the data points by joining them by straight lines
(trapezoidal rule). In the low energy region it is customary to
present the $\pi^+ \pi^-$ and $\rho \ra \pi^+ \pi^-$ data in terms of
the absolute square of the pion form-factor $|F_\pi(s)|^2$ which is
related to the cross section by
\be
\sigma(\epm \ra \pi^+ \pi^-)=
\frac{\pi}{3}\frac{\al^2\beta_\pi^3}{s}|F_\pi(s)|^2\;\;\;
\mbox{or}\;\;\;\;\; R_{\pi\pi}=\frac{\beta_\pi^3}{4}|F_\pi(s)|^2
\ee
where $\beta_\pi=(1-4m^2_\pi/s)^{1/2}$ is the pion velocity. In
contrast to previous works which were using Gounaris-Sakurai
\ci{gusa} kind of parametrizations we will rely on the direct
integration of the data in order not to obscure the error estimates.
The definition of the pion form factor is similar in spirit to the one
for $R(s)$. In principle, it is defined as a ratio of hadronic to
leptonic cross sections such that it corresponds to a purely hadronic
matrix element. For most of the low energy experiments (in particular
the Novosibirsk VEPP-2M ones) $R$ was calculated as follows
\ci{Eidelman78}: first the hadronic cross section was obtained.
Radiative corrections were applied which included lepton vacuum
polarization only. And then the corrected hadronic cross section was
divided by the muon cross section calculated using the constant low
energy $\alpha$.  The more precise accounting of using the running
$\alpha$ with the inclusion of hadronic vacuum polarization leads to a
numerically very small ($a_\mu$) or negligible ($\dal$) difference.
For consistency we apply the factor $(1+2 \Delta \al_l (s))(\al / \al
(s))^2$ to the pion form factor to account for this missing
correction. The remarks made here for $\pi^+
\pi^-$ apply as well for the $K^+ K^-$ and $K_S K_L$ form-factors.

\subsection{Resonances}
\label{sec:RESO}

A few exceptions from the direct integration are the narrow resonances
$\omega$, $\phi$, the $J/\psi$ family (6 states) and the $\Upsilon$
family (6 states). Here we can safely use the parametrization as
Breit-Wigner resonances
\be
\sigma_{BW}(s)=\frac{12 \pi}{M_R^2}\frac{\Gamma_{ee}}{\Gamma_R}
\frac{M_R^2 \Gamma_R \Gamma(s) }{(s-M_R^2)^2+M_R^2 \Gamma^2(s)}
\label{BWR}
\ee
or as a zero width resonance
\be
\sigma_{NW}(s)=\frac{12 \pi^2}{M_R} \Gamma_{ee}\: \delta(s-M_R^2)
\label{NWA}
\ee
using resonance parameters from the Review of Particle
Properties~\ci{PDG}.  The $s$--dependent width is defined by the
imaginary part of the irreducible self--energy of the resonance
propagator as ${\rm Im}
\Pi(s)=M_R \Gamma(s)$. Sufficiently far above the thresholds it behaves
like
\be
\Gamma(s) \simeq s/M_R^2\:\Gamma_R \;\;\;{\rm where}\;\;\;\;\;
\Gamma_R=\Gamma(M_R^2)\;\;.
\ee

The procedure of calculating widths for the narrow resonances of the
$J/\psi$ and $\Upsilon$ families is described in the Review of
Particle Properties, e.g., in the 1994 edition, on page 1661. The
particle data group (PDG) lists ``dressed''(i.e. physical) widths
$(\Gamma, \Gamma_{ee},\cdots)$ rather than lowest order ones
$(\Gamma^{(0)}, \Gamma_{ee}^{(0)},\cdots)$. This convention exists
since the 1988 edition of the Review of Particle
Properties~\ci{PDG88}, and was triggered by the articles
Refs.~\ci{Buchmueller88}, \ci{Koenigsmann87} and
\ci{Alexander88} which pointed out that, in the past, different
experiments have been using different conventions which depended on
the different treatment of the radiative corrections. A discussion of
the radiative corrections which were performed by different
experiments for the $\Upsilon$ resonances may be found in
Ref.~\ci{Buchmueller88} where consistent world averages for the
dressed widths are calculated by appropriate rescaling of the peaks of
the resonances. In Ref.~\ci{Alexander88} the $J/\psi$ and the
$\Upsilon$ resonances were refitted using state of the art
calculations for resonance line-shapes as they have been developed for
precision physics at the $Z$-resonance, using Eq.~\re{BWR} with
$\Gamma(s)=s/M_R^2\:\Gamma_R $. Note that the convention mentioned
above is employed by the authors themselves. The PDG only lists the
results of original papers and averages them.

In terms of the physical (``dressed'') widths resonance contributions
to the $R$--values are given by
\be
R_{BW}=\frac{9}{\alpha^2(s)}
\frac{s}{M_R^2}\frac{\Gamma_{ee}}{\Gamma_R}
\frac{M_R^2 \Gamma_R \Gamma(s)}{(s-M_R^2)^2+M_R^2 \Gamma^2(s)}
\ee
and correspondingly for the narrow resonance approximation.  The
latter is used only for calculating the resonance contributions to
$a_\ell$. It should be noted that the Breit--Wigner formula itself,
valid in the vicinity of the resonance, is a result of summing over
all quark vacuum polarization diagrams with any number of loops.

For the $\omega$ and the $\phi$ we proceed as described in
Ref.~\ci{Aulchenko87} (see also \ci{Achasov76}) and use the
relativistic Breit--Wigner form with a $s$--dependent width
\ba
 \Gamma_\omega (s) &=& \Gamma (\omega \ra 3 \pi,s) +\Gamma (\omega \ra
\pi^0 \gamma,s) +\Gamma (\omega \ra 2 \pi,s) \crn &=&
\frac{s}{M_\omega^2}\Gamma_\omega \left\{~~ Br(\omega \ra 3
\pi)\frac{F_{3\pi}(s)}{F_{3\pi}(M^2_\omega)} \right. \crn & &
~~~~~~~~~~ \left.  + Br(\omega \ra \pi^0 \gamma)
\frac{F_{\pi \gamma}(s)}{F_{\pi \gamma}(M^2_\omega)}
+ Br (\omega \ra 2 \pi)\frac{F_{2\pi}(s)}{F_{2\pi}(M^2_\omega)}
\right\} \crn
&& \crn \Gamma_\phi (s) &=& \Gamma (\phi \ra K^+ K^-,s) +\Gamma (\phi
\ra K_S K_L,s) +\Gamma (\phi \ra 3 \pi,s) +\Gamma (\phi \ra \pi^0
\gamma,s) +\Gamma (\phi \ra \eta \gamma,s) \crn &=&
\frac{s}{M_\phi^2}\Gamma_\phi \left\{~~ Br(\phi \ra K^+ K^-)\frac{F_{
K^+ K^-}(s)}{F_{ K^+ K^- }(M^2_\phi)} + Br(\phi \ra K_S K_L)\frac{F_{
K_S K_L}(s)}{F_{ K_S K_L }(M^2_\phi)}
\right. \crn & &~~~~~~~~~~ \left.
+ Br (\phi \ra 3 \pi)\frac{F_{3\pi}(s)}{F_{3\pi}(M^2_\phi)}
\right. \crn & &~~~~~~~~~~ \left.
+ Br(\phi \ra \pi^0 \gamma)\frac{F_{\pi \gamma}(s)}{F_{\pi
\gamma}(M^2_\phi)} + Br(\phi \ra \eta \gamma)\frac{F_{\eta
\gamma}(s)}{F_{\eta \gamma}(M^2_\phi)}
\right\}
\ea
where $Br(V \ra X )$ denotes the branching fraction for the channel
$X$ and $F_{X}(s)$ is the phase space function for the corresponding
channel normalized such that $F_{X}(s) \ra const$ for $s \ra \infty$.
For the two-body decays $V \ra P_1 P_2$ we have $F_{P_1
P_2}(s)=(1-(m_1+m_2)^2/s)^{3/2}$. The channel $V \ra 3 \pi$ is
dominated by $V \ra \rho \pi \ra 3\pi$ and this fact is used when
calculating $F_{3\pi}(s)$~\ci{Achasov76}.  Before extracting the
width, radiative corrections according to Bonneau and
Martin~\ci{BonneauMartin} or Kuraev and Fadin~\ci{Eidelman78} have
been performed which include subtracting the electron contribution to
the vacuum polarization.  In these cases the correction to be applied
is $(1+2 \Delta \al_e (s))(\al / \al (s))^2$ and not the full one.

\subsection{The $\pi^+ \pi^-$ threshold region}
\label{sec:CHPT}

Experimental data are poor below about 400 MeV because the cross
section is suppressed near the threshold. Because of the $1/s^2$
weight factor for small $s$ in Eq.~\re{amu} we have to worry whether
there could not be a relevant contribution missing. Here results from
chiral expansion of the pion form factor \ci{Juerg} can be used (see
also Ref.~\ci{CLY}).

To a good approximation the relevant vector form factor is given by
\be
F_V^{CHPT}\simeq 1+\frac16 <r^2>^\pi_V \cdot s\:+\:c^\pi_V \cdot s^2
\label{chpt1}
\ee
with $<r^2>^\pi_V$=0.427$\pm$0.010 fm$^2$ and
$c^\pi_V$=4.1$^{+0.2}_{-0.6}$ GeV$^{-4}$.  The pion charge radius used
here was determined from the precise spacelike data in
Ref.~\ci{Amendolia86}.  We have used the value obtained from fits with
a free normalization.  The error includes the 0.9\% systematic error
of the data. The crucial point here is that the threshold behavior is
severely constrained by the chiral structure of QCD via the rather
precise data for the pion form factor in the spacelike region. The
convergence of the momentum expansion can be improved by using the
Pad\'{e} approximants rather than the asymptotic expansion itself
which is denoted by [0,2] in Pad\'{e} terminology.
\begin{figure}[hb]
\vspace*{7.4cm}
\epsffile[-100 0 75 50]{chpt.pos}
\end{figure}
\begin{center}
\begin{minipage}[h]{12.3cm} \baselineskip 10truept
\begin{tabular}{ll}
{\bf Fig.~2:} &
\begin{tabular}[t]{p{100mm}}
The $\epm \ra \pi^+ \pi^-$ data near threshold compared with the
prediction of the pion form factor $|F_\pi(s)|^2$ for timelike $s=E^2$
from chiral expansion to two-loop order \ci{Juerg}.  [1,1] and [2,0]
denote Pad\'{e} improvements.
\end{tabular} \end{tabular}
\end{minipage}
\end{center}
With the information we have on the expansion coefficients we may
obtain the [2,0] form
\be
F_V^{CHPT}\simeq 1/(1-c_1 \cdot s\:-\:(c_2-c_1^2)\: \cdot s^2)\;\;,
\ee
with $c_1=\frac16 <r^2>^\pi_V$ and $c_2=c^\pi_V$; or the [1,1] form
\be
F_V^{CHPT}\simeq (1+\:(c_1-c_2/c_1)\:\cdot s)/(1-\:(c_2/c_1)\: \cdot
s)\;\;,
\ee
which agree with \re{chpt1} when expanded in $s$ up to terms of
unknown higher orders. The difference obtained from the different
representations is the ``model'' error, uncertainties due to missing
higher order terms.

The results are shown in Fig.~2 and provide a good description of the
data in the timelike region.

\subsection{Estimate of the error}
\label{sec:errorest}

While statistical errors are added in quadrature throughout in our
analysis the systematic errors of an experiment have to be added
linearly. Usually the experiments give systematic errors as a relative
systematic uncertainty and the systematic error to be added linearly
is given by the central value times the relative uncertainty. For data
from different experiments the combination of the systematic errors is
more problematic. If one would add systematic errors linearly
everywhere, the error would be obviously overestimated since one would
not take into account the fact that independent experiments have been
performed. Since we are interested in the integral over the data only,
a natural procedure seems to be the following: for a given energy
range (scan region) we integrate the data points for each individual
experiment and then take a weighted mean, based on the quadratically
combined statistical and systematic error, of the experiments which
have been performed in this energy range.  By doing so we have assumed
that different experiments have independent systematic errors, which
of course often is only partially true\footnote{If there are known
common errors, like the normalization errors for experiments performed
at the same facility, one has to add the common error after averaging.
In some cases we correct for possible common errors by scaling up the
systematic error appropriately.}. The problem with this method is that
there exist regions where data are sparse yet the cross section varies
rapidly, like in the $\rho$-resonance region.  The applicability of
the trapezoidal rule is then not reliable, but taking other models for
the extrapolation introduces another source of systematic errors. It
was noticed some time ago in Ref.~\ci{pitfal} that fitting data to
some function by minimizing $\chi^2$ may lead to misleading
results\footnote{The problem addressed in Ref.~\ci{pitfal} is that
``The best fits to the data which are affected by systematic
uncertainties on the normalization factor have the tendency to produce
curves lower than expected, if the covariance matrix of the data
points is used in the definition of $\chi^2$''.} and we insist on
avoiding this kind of problems.

In order to start from a better defined integrand we do better to
combine all available data points into a single dataset. If we would
take just the collection of points as if they were from {\em one}
experiment we not only would get a too pessimistic error estimate but
a serious problem could be that scarcely distributed precise data
points do not get the appropriate weight relative to densely spaced
data point with larger errors. What seems to be more adequate is to
take for each point of the combined set the weighted average of the
given point and the linearly interpolated points of the other
experiments:
\bea
\bar{R}=\frac{1}{w}\sum_i w_i R_i
\eea
with total error $\delta_{tot}=1/\sqrt{w}$, where $w=\sum_i w_i$ and
$w_i=1/\delta_{i\;tot}^2\:$. By
$\delta_{i\;tot}=\sqrt{\delta_{i\;sta}^2+\delta_{i\;sys}^2}$ we denote
the combined error of the individual measurements. In addition, to
each point a statistical and a systematic error is assigned by taking
weighted averages of the squared errors:
\bea
\delta_{sta}=\left( \frac{1}{w}\sum_i w_i\: \delta_{i\;sta}^2
\right)^{1/2}
\;\;,\;\;\;
\delta_{sys}=\left( \frac{1}{w}\sum_i w_i\: \delta_{i\;sys}^2
\right)^{1/2}\;\;.
\eea
There is of course an ambiguity in separating the well--defined
combined error into a statistical and a systematic one. We may also 
calculate separately the total error and the statistical one and
obtain a systematic error
$\delta_{sys}=\sqrt{\delta_{tot}^2-\delta_{sta}^2}$. Both procedures
give very similar results. We also calculate $\chi^2=\sum_i
w_i\:(R_i-\bar{R})^2$ and compare it with $N-1$, where $N$ is the
number of experiments.  Whenever $S=\sqrt{\chi^2/(N-1)}>1\:,$ we scale
the errors by the factor $S$, unless there are plausible arguments
which allow one to discard inconsistent data points.

\section{The $\epm$ data and the origin of systematic errors}

Some general comments concerning the $R$ determination are in order.
In the ideal case one directly identifies each possible annihilation
channel and measures its cross section. After that $R$ is obtained as
a sum of all separate contributions. We will call such an approach an
``exclusive'' one.  One should be careful, however, while estimating
the resulting systematic uncertainty since systematic uncertainties of
separate channels may contain common parts like, e.g., normalization
uncertainties. There may also be some model uncertainties arising from
the fact that for each specific channel one has to assume some
specific mechanism of the production of final particles, for example,
for the production of four pions, $\omega \pi$, $a_1 \pi$, $\rho
\pi \pi$ etc., while in reality particles can be produced by different
mechanisms. In a more realistic case identification of separate
channels is not possible. One observes a certain number of
multi-hadronic events of different types, for example, 2 charged
particles, 2 charged particles plus $n$ photons, 3 charged particles
etc., assumes some mechanism of multi-hadron production and calculates
within it a detection efficiency of observing any configuration of
charged particles and photons. Minimizing $\chi^2$ one obtains then
the cross sections of separate channels and their sum gives a total
cross section. In old experiments, pion production with an invariant
phase space distribution was used as a mechanism of particle
production (see, e.g.~\ci{Bacci79}). Lately the LUND model \ci{LUND}
has been used for detection efficiency determination. Here again
additional model uncertainties arise. The value of $R$ determined by
such an ``inclusive'' method usually yields the total cross section
not including production of two-body final states. One should also
take into account that there may be production channels ``invisible''
to some experiments, like, for example, those having only neutral
particles in the final state. Some of them may be accounted for by
using isospin symmetry. For example, the cross section of the
reaction $e^+e^-
\ra
\pi^+\pi^-3\pi^0$ is equal to one half of that for $e^+e^- \ra
2\pi^+2\pi^-\pi^0$, independently of the production mechanism. Such
corrections although small will influence the central value of the
calculated integral and will also contribute to the systematic
uncertainty.

    Since 1985 when the paper of Kinoshita et al. \ci{Kino2} was
published, a lot of new experimental data on the $R$ measurement in
$\epm$ annihilation into hadrons has been accumulated. Progress in the
low energy range was mostly due to experiments at VEPP-2M at
Novosibirsk where three groups (OLYA, ND and CMD) provided
independently the information on different exclusive channels of
$\epm$ annihilation and at DCI at Orsay coming from the DM2
experiment. In the Novosibirsk experiments the center of mass
(c.m.) energy range from the threshold of hadron production up to 1.4
GeV was studied, whereas Orsay experiments covered the c.m. energy
range from 1.35 up to 2.3 GeV. The high integrated luminosity
collected in these experiments allowed them to improve considerably their
statistical precision. Larger solid angles of the detectors and the use
of electromagnetic calorimeters providing detection of photons with
good energy resolution facilitated the identification of the large
number of exclusive annihilation channels.

     In the Novosibirsk energy range the cross section is dominated by
$\rho$, $\omega$ and $\phi$ resonances, resulting in its strong
energy dependence.  Typical multiplicities in multi-pion production do
not exceed five, making the total number of accessible channels rather
small. Besides that, the cross section of the two-body channels is
dominating the total cross section.  All these circumstances make the
usual inclusive procedure of $R$ determination from the total number
of multi-hadronic events of a different type almost meaningless and
subject to large uncertainties. Therefore experimental efforts were
aimed at the selection of exclusive channels followed by the
``exclusive'' $R$ determination by simply adding contributions of
different reactions. The energy range from 1.4 up to 2.3 GeV
systematically studied by Orsay and Frascati groups is much worse
understood. It is clear that higher vector mesons play an important
role, but their precise parameters are not yet known, thus making a
phenomenological model approach rather ambiguous (see the recent paper
\ci{Clegg94}). The cross section of multi-hadronic channels is
considerably larger than that of two-body channels. However, since the
multiplicity and therefore the number of possible annihilation
channels is still not too big, the approach can be twofold: studies of
exclusive channels can be complemented by the independent
``inclusive'' $R$ determination. At larger energies (above
approximately 2 GeV) data on the cross sections of separate channels
are missing, only few exclusive channels have been measured (usually
in the near resonance region) and there are ``inclusive'' $R$
determinations only.

     The reaction $\epm \ra \pi^+\pi^-$ was studied with high
precision by the two groups OLYA and CMD which found good agreement
between each other and the results of the joint analysis were
published in Ref.~\ci{Barkov85}. In the CMD experiment 24 points from
360 to 820 MeV have been studied with a systematic uncertainty less
than 2\%. OLYA performed scanning of the energy region from 640 to
1400 MeV with a small energy step and had a systematic uncertainty
from about 4\% at the $\rho$-meson peak up to 15\% at 1400 MeV.  Also
used were the older data near threshold from OLYA~\ci{Vasserman79},
TOF~\ci{Vasserman81}, NA7~\ci{Amendolia84} and VEPP-2M~\ci{Barkov79}
as well as the measurements from 483 to 1096 MeV by
DM1~\ci{Quenzer78}. The data are shown in Fig.~3. At higher energies
this reaction was systematically studied by DM2~\ci{Bisello89}. Older
results from $\mu
\pi$~\ci{Barbiellini73}, BCF~\ci{Bollini75}, and MEA~\ci{Esposito80}
at Frascati are also included in the analysis.

     The reaction $\epm \ra \pi^+\pi^-\pi^0$ was studied by different
Orsay and Novosibirsk groups at the $\omega$ and $\phi$ meson (for
these data we used the values of the leptonic widths from the Review
of Particle Properties~\ci{PDG} to calculate analytically the resonance
contributions) as well as in the off--resonance region by
CMD~\ci{Barkov89}, ND~\ci{Dolinsky91} at Novosibirsk and
M2N~\ci{Parrour76}, M3N~\ci{Cosme79}, DM1~\ci{Cordier80}, and
DM2~\ci{Antonelli92} at Orsay.

      There are two channels in four-pion production:
$\pi^+\pi^-\pi^0\pi^0$ and $2\pi^+2\pi^-$.
OLYA~\ci{Kurdadze86,Kurdadze88} and ND~\ci{Dolinsky91} who scanned the
energy region from about 640 to 1400 MeV provided results on both,
while CMD~\ci{Barkov88} measured the cross section of the latter in 9
points from 1019 up to 1403 MeV. The values of the cross section
determined by ND are usually higher than those of OLYA in both
reaction channels; however, they are within systematic uncertainties
which are estimated by the authors to be 15\% and 10\% respectively
for ND and 20\% in both cases for OLYA.  CMD claimed a 10\% systematic
uncertainty and within it agreed with both ND and OLYA. At higher
energies we used the data from M3N~\ci{Cosme79}, MEA~\ci{Espos80},
DM1~\ci{Cordier82} and DM2~\ci{Schioppa86,Bisello91}. Since we are
interested in the values of total cross sections only, the mechanism
of particle production is in general not important for our analysis.
However, one should note that the reaction $\epm \ra
\omega \pi^0$ plays an important role in the production of the
$\pi^+\pi^-2\pi^0$ final state. Since $\omega$ has a 8.5\% branching
ratio for the $\pi^0 \gamma$ decay \ci{PDG} it should be taken
into account separately since it will be missed in the inclusive
analysis. Accordingly we add $\sigma_{\omega \pi^0} B(\omega \ra \pi^0
\gamma)$ to the total cross section, a contribution which was
previously ignored.

      The cross section of the reaction $\epm \ra 2\pi^+2\pi^-\pi^0$
was measured at low energy by CMD~\ci{Barkov88} and at higher energy
by different groups. The most precise of them are those by
DM1~\ci{Cordier81} and DM2~\ci{Antonelli92}. Its isospin partner $\epm
\ra \pi^+\pi^-3\pi^0$ is much worse studied, but as mentioned above,
the rigorous isotopic symmetry relation requires that its cross
section be two times smaller. Since $\omega \pi \pi$ is a dominant
mechanism of five pion production, one should make a correction for it
similar to that for the reaction $\epm \ra \omega \pi^0$.

      Six pion production is much less studied. There are three
possible final states of which only $3\pi^+3\pi^-$ and
$2\pi^+2\pi^-2\pi^0$ have been observed.  The isotopic symmetry does
not give unfortunately the rigorous relations between different
isotopic partners~\ci{Pais60}; however, one can expect that the
corrections for the missing parts are small. The existing data are
rather controversial.  The $\gamma \gamma 2$~\ci{Bacci79} measurements
are typically characterized by much higher values of the cross section
hardly compatible with the measurements of DM1~\ci{Bisello81} and
DM2~\ci{Schioppa86}. Measurements at low energy are also available
from CMD~\ci{Barkov88}.

      The reaction $\epm \ra \eta \pi^+\pi^-$ was studied by
\ci{Antonelli88} and \ci{Dolinsky86}.

      Production of kaon pairs was studied by
OLYA~\ci{Ivanov81,Ivanov82}, CMD~\ci{Anikin83}, DM1~\ci{Mane81} and
DM2~\ci{Bisello88}.

      DM1~\ci{KKpi1} and DM2~\ci{Schioppa86} measured also cross
sections of the reactions $\epm \ra K\bar{K} \pi$ and
$K\bar{K}\pi\pi$.  However, not all possible combinations have been
studied. The corrections for the missing modes should be applied.

      Information on the production of baryons is scarce. Measurements
by DM1~\ci{Delcourt79}, DM2~\ci{Bisello83} and FENICE~\ci{Antonelli94}
showed that the cross section is small. However, it should be taken
into account if $R$ is determined inclusively.

Our compilation of $R$, obtained following the procedure described in
Sec.~\ref{sec:errorest}, is shown in Figs.~4 to 8, together with
original compilations by the experimental groups themselves. It should
be mentioned that part of the large systematic errors given by
experiments in particular below the $J/\psi$ are supposed to account
for the lack of understanding or performing the radiative corrections.
Most experiments have applied radiative corrections including at least
the vacuum polarization contribution from the electron, which is the
dominant contribution at low
energies~\ci{BonneauMartin,BerendsKleiss,Eidelman78}. The hadronic
contribution to the vacuum polarization has been known since the work
of Berends and Komen~\ci{Berends76} in 1976 but usually it was not
included in the QED corrections. We shall assume that on the average
experiments have subtracted only the electron contribution to the
vacuum polarization and accordingly rescale the $R$-values by $(1+2
\Delta \al_e)(\alpha/\alpha(s))^2$ below the $J/\psi$. As we mentioned
before this remark applies to the $\omega$ and $\phi$ resonances as
well.

In the high energy region we distinguish the $J/\psi$ and the
$\Upsilon$ resonances and the background inclusive measurements of the
total hadronic cross section which is usually presented in terms of
$R$-values.  The resonance contributions are taken into account as
explained in Sec.~\ref{sec:RESO} Masses, widths and the electronic
branching fractions are taken from the Review of Particle
Properties~\ci{PDG}.  Since only total errors are given for these
quantities, we treat the error from the mass (negligible) and width
quadratically and the one from the branching fraction linearly within a
family. In this way we take into account that the systematic errors
should be added linearly.

In the region from the $J/\psi$ to the $\Upsilon$ $R$-measurements are
available from Mark~I~\ci{Bacino78,Siegrist82}, DASP~\ci{DASP},
PLUTO~\ci{Criegee82}, LENA~\ci{Niczyporuk82}, Crystal Ball
(CB)~\ci{Edwards90} and MD-1~\ci{Blinov91} (see Fig.~7). The Crystal Ball
Collaboration has carefully reanalyzed their old data and obtained
$R(s)$ values substantially lower than Mark I and in agreement with
the other experiments PLUTO, LENA and MD-1. The results are now much
closer to expectations from perturbative QCD ($R\sim 10/3$ at lowest
order). The change of the data is mainly due to a up-to-date treatment
of the QED radiative corrections and $\tau$ subtraction.
Ref.~\ci{Edwards90} gives a detailed description of how precisely the
radiative corrections have been performed. The procedure is based on
the calculation of Ref.~\ci{BerendsKleiss} which has been applied by
all DESY experiments. Thus, except from the Mark I data, all data above
the $J/\psi$ resonance have published properly normalized
$R$--values.  The Mark I values we rescale by $(1+2 \Delta \al_l)
(\alpha/\alpha(s))^2$, because we assume that hadronic vacuum
polarization effects have not been subtracted, while the leptonic
correction has been taken into account in a non-resummed form.

Below we will study the option of including and discarding the Mark I
data in the overlapping range, where the Mark I data systematically
lie 28\% higher as we can see in Fig.~7.

Above the $\Upsilon$ we include PETRA and PEP data up to $E_{\rm cut}=40$
GeV. In this range we correct for the $\gamma-Z$ mixing contribution
as described earlier. For larger energies the $\gamma-Z$ mixing would
be substantial and make the analysis less transparent. In fact
perturbative QCD is reliable for evaluating the high energy
contribution already at lower energies. In the PETRA/PEP range there
are many measurements of rather high accuracy available, a fact which
is particularly important for the precise determination of
$\alpha(M_Z)$.
Data are mainly from CELLO~\ci{CELLO}, JADE~\ci{JADE},
MARK~J~\ci{MARKJ} and TASSO~\ci{TASSO}.  Further included are the
$R$-measurements from DASP~\ci{DASP}, DHHM~\ci{DHHM}, CLEO~\ci{CLEO},
CUSB~\ci{CUSB}, HRS~\ci{HRS}, MAC~\ci{MAC} and MD-1~\ci{Blinov91}(see
Fig.~8).

\section{Results}

The results presented here have been obtained following the procedure
described in Sec.~\ref{sec:errorest} For the $\pi^+\pi^-$ data we have
three sets of data points as collected in Ref.~\ci{Barkov85}. They are
displayed in Fig.~3 and we label them by DM1$^+$ (which includes data
from DM1, NA7 and TOF), CMD and OLYA. In the range 0.81-1.4 GeV we use
a compilation similar to the one in Ref.~\ci{Dolinsky91} but with all
available data which are shown in Fig.~4.  This is done by adding up
the individual channels, where for each channel weighted averages are
taken as described in Sec.\ref{sec:errorest} The $\phi$-resonance is
taken into account in analytic form in the narrow interval between
1.00 and 1.04 GeV. Outside this interval the $\phi$ contribution is
included in the background. The cut at 1.4 GeV is justified as it is
the energy limit of VEPP-2M. From 1.4 to 2.3 GeV we combine
$R$--values published by the experiments MEA, $\gamma\gamma 2$,
$\mbox{B}\bar{\mbox{B}}$, M3N, DM1 and DM2; from 2.3 to 3.1 GeV data
collected by $\gamma\gamma 2$, DM2 and Mark I are available, where the
last two give $R$ while the others $R(n>2)$, in which case the 2-body
channels have to be taken into account separately. In taking weighted
averages here we did not account yet for the fact that some
measurements have been performed at the same machine and hence have
common normalization errors. We therefore enlarge the systematic error
in this domain by a factor $\sqrt{2}$ to be on the conservative side.
The data for this region are shown in Figs.~5 and 6.  In the region
between the $J/\psi$ and $\Upsilon$ resonances we again compare two
methods. First we combine data by calculating the weighted averages as
described above. The Mark I data are treated as an independent set,
and are eventually combined after integration.  Alternatively, we
split the region in such a way that results from different experiments
can be combined after integration.  From 3.1 to 3.6 GeV we have data
from Mark I, form 3.6 to 5.2 GeV Mark I, DASP and PLUTO, from 5.2 to
7.2 GeV Mark I, CB and PLUTO and from 7.2 to 9.46 GeV PLUTO and MD-1.
Here we take weighted averages after integration. The different energy
ranges mentioned are treated as independent when adding up the
results, a procedure which again may not be fully justified.

The data above the $J/\psi$ still include the resonance contributions
$\psi(4040)$, $\psi(4160)$ and $\psi(4415)$, which we subtract and
include in the $J/\psi$--family resonance contribution.  Finally from
the $\Upsilon$ to 40 GeV we take weighted averages for the data sets
JADE, TASSO, MD-1 and others as shown in Fig.~8.  Consistent results
are obtained if we combine the data pointwise and add all systematic
errors linearly. The results for $a_{\mu}^{{\rm had}}$ are presented
in Table~3. The last two
\bet{|c|c|r|c|c|}
\multicolumn{5}{c}{Table~3a: Contributions to
$a_{\mu}^{{\rm had}}\cdot 10^{10}$}\\
\hline
\hline
 final state &energy range (GeV) &contribution (stat) (syst)&
 rel. err.& abs. err. \\
\hline
  $ \rho $   &(0.28, 0.81)    &426.66 ( 5.61) (10.62)& 2.8\% & 1.7\% \\
  $ \omega $ &(0.42, 0.81)    & 37.76 ( 0.45) ( 1.02)& 3.0\% & 0.2\% \\
  $ \phi   $ &(1.00, 1.04)    & 38.55 ( 0.54) ( 0.89)& 2.7\% & 0.1\% \\
  $ J/\psi $ &                &  8.60 ( 0.41) ( 0.40)& 6.7\% & 0.1\% \\
  $\Upsilon$ &                &  0.10 ( 0.00) ( 0.01)& 6.7\% & 0.0\% \\
hadrons      &(0.81, 1.40)    &112.85 ( 1.33) ( 5.49)& 5.0\% & 0.8\% \\
hadrons      &(1.40, 3.10)    & 56.43 ( 0.45) ( 7.22)&12.8\% & 1.0\% \\
hadrons      &(3.10, 3.60)    &  4.47 ( 0.23) ( 0.86)&19.9\% & 0.1\% \\
hadrons      &(3.60, 9.46)    & 14.06 ( 0.07) ( 0.90)& 6.5\% & 0.1\% \\
hadrons      &(9.46, 40.0)    &  2.70 ( 0.03) ( 0.13)& 4.9\% & 0.0\% \\
perturbative &(40.0, $\infty$ ) &  0.16 ( 0.00) ( 0.00)& 0.2\% & 0.0\% \\
\hline
\hline
   total     &                &702.35 ( 5.85) (14.09)& 2.2 \% & 2.2\% \\
\hline
\hline
\ent
columns give the relative uncertainty (rel.err.) of the individual
contribution and the absolute uncertainty (abs.err.) relative to the
total result.  The $\rho$-contribution always includes a contribution
2.08 (0.01)(0.05) calculated using chiral perturbation theory from the
2$\pi$-threshold to 318 MeV where data points start. For details we
refer to Sec.~\ref{sec:CHPT} In spite of a major effort in the $\rho$
region the result is almost the same as the one obtained using
Kinoshita's fit~\ci{Kino2} which was used subsequently in
\ci{jegup,Jegerlehner86,Burkhardt89,TASI}. A previous analysis
\ci{jegup} gave the results 428.95(1.71)(12.81)[426.68(1.70)(12.74)]
for the $\rho$ contribution in the range (0.28, 0.81) GeV and
125.54(3.89)(11.61)[124.27(3.83)(11.46)] for the range (0.81,1.40)
GeV. The values in square brackets are the ones obtained after
correcting for the missing subtraction of the hadronic vacuum
polarization. While the $\rho$ contribution below 0.81 GeV remains
unchanged the $\rho$--tail plus background up to 1.4 GeV turns out to
be somewhat smaller.

If we would treat the different experiments separately and then take
weighted averages in common domains we would obtain 436.30(9.95)(8.02)
for the $\rho$ up to 0.81 GeV, a somewhat higher value. Details are
illustrated by Tab.~3b. The high value is the result of a failure of
\bet{c|c|c|c}
\multicolumn{4}{c}{Table~3b: $\rho$--contributions to
$a_{\mu}^{{\rm had}}\cdot 10^{10}$ of different experiments}\\
\hline
 Energy range: & (0.32, 0.36) GeV&(0.36, 0.40) GeV&(0.40, 0.81) GeV\\
\hline
 ~DM1$^+$ & 6.213 (0.372)(0.214)& 9.225 (0.432)(0.197)&
409.739(14.450)( 9.331)\\
CMD & -- & 11.533 (0.710)(0.231)&
414.176(11.051)( 8.284)\\
OLYA & -- & -- &
472.399(24.364)(22.672)\\
\hline
Averaged & 6.213 (0.372)(0.214)& 9.890 (0.373)(0.146)&
418.120( 9.939)( 7.661)\\
\hline
Tab.~3a & 6.177 (0.370)(0.210)& 9.363 (0.582)(0.269)&
409.034( 5.572)(10.136)\\
\hline
\ent
the trapezoidal rule when applied to the OLYA data. The reason is that
the three lowest points have much larger energy separations than the
other points at higher energies and a high weight due to the $1/s^2$
behavior of the kernel. Therefore the first three points yield a
large contribution to the integral. Actually, applying the trapezoidal
rule here we overestimate the contribution because the low energy tail of
the resonance is a strongly varying concave function. This problem is
largely circumvented by combining data from different experiments
before integration.
Note that, while the integrals taken over the whole range agree quite
well between different experiments, the individual contributions in
the subdomains do not agree within errors.  Apparently the integral is
better defined than the local values. In other words, the $\rho \ra
\pi\pi$ region is after all not as well established as the
experiments claim. There are many points where the weighted average
yields $S>1$ and the error must be enlarged.

If we just take the collection of all points in a given energy region
and add systematic errors linearly we obtain 413.80(8.10)(23.81) for
the range up to 0.81 GeV, with a much larger systematic error.
Similarly, we find 113.71(1.31)(12.31) for the contribution from 0.81
to 1.40 GeV. The final result would be 695.42(8.30)(27.52).  We
remind the reader that in the important region from 0.81 to 1.4 GeV
data for the channels $\pi^+ \pi^- 2\pi^0$ and $\pi^+ \pi^-\pi^+
\pi^-$ from ND lie substantially higher than from other experiments as
can be seen in Fig.~4. A new experiment is required to
resolve the current discrepancy.

In Tab.~3c we give the results we obtain when using the resummed
photon propagator in the diagram Fig.~1. This dressed contribution we
denote by $a_{\mu}^{{\rm had}\:*}$. We notice that the difference
between the dressed and the undressed form is about 23, slightly
larger than the uncertainty of 16 and of the same size as the
interesting weak contribution $a_\mu^{\rm weak}\simeq 20\:.$ For
$a_\tau$ the results are collected in Tab.~4.  The hadronic
contribution to the electron anomaly is $a_e^{{\rm had}\:*}=(194.48
\pm 1.69 \pm 3.87) \times 10^{-14}$.

\bet{|c|c|r|}
\multicolumn{3}{c}{Table~3c: Contributions to
$a_{\mu}^{{\rm had}\:*}\cdot 10^{10}$}\\
\hline
\hline
 final state &energy range (GeV) &contribution (stat) (syst)\\
\hline
  $ \rho $   &(0.28, 0.81)    &438.63 ( 5.76) (10.91)\\
  $ \omega $ &(0.42, 0.81)    & 38.90 ( 0.47) ( 1.05)\\
  $ \phi $   &(1.00, 1.04)    & 39.86 ( 0.56) ( 0.92)\\
  $ J/\psi $ &                &  9.05 ( 0.43) ( 0.42)\\
  $\Upsilon$ &                &  0.11 ( 0.00) ( 0.01)\\
hadrons      &(0.81, 1.40)    &116.84 ( 1.38) ( 5.69)\\
hadrons      &(1.40, 3.10)    & 58.92 ( 0.47) ( 7.55)\\
hadrons      &(3.10, 3.60)    &  4.71 ( 0.25) ( 0.90)\\
hadrons      &(3.60, 9.46)    & 14.91 ( 0.08) ( 0.96)\\
hadrons      &(9.46, 40.0)    &  2.94 ( 0.03) ( 0.14)\\
perturbative &(40.0, $\infty$ ) &  0.18 ( 0.00) ( 0.00)\\
\hline
\hline
   total     &                &725.04 ( 6.01) (14.57)\\
\hline
\hline
\ent

\bet{|c|c|r|}
\multicolumn{3}{c}{Table~4: Contributions to
$a_{\tau}^{{\rm had}\:*}\cdot 10^{8}$}\\
\hline
\hline
 final state &energy range (GeV) &contribution (stat) (syst)\\
\hline
  $ \rho $   &(0.28, 0.81)    &141.35 (1.66) (3.42)\\
  $ \omega $ &(0.42, 0.81)    & 15.06 (0.18) (0.41)\\
  $ \phi $   &(1.00, 1.04)    & 21.06 (0.29) (0.48)\\
  $ J/\psi $ &                & 13.47 (0.64) (0.65)\\
  $\Upsilon$ &                &  0.26 (0.01) (0.01)\\
hadrons      &(0.81, 1.40)    & 58.23 (0.64) (3.11)\\
hadrons      &(1.40, 3.10)    & 58.84 (0.56) (7.94)\\
hadrons      &(3.10, 3.60)    &  7.02 (0.36) (1.35)\\
hadrons      &(3.60, 9.46)    & 28.04 (0.14) (1.80)\\
hadrons      &(9.46, 40.0)    &  7.36 (0.07) (0.35)\\
perturbative &(40.0, $\infty$ ) &  0.50 (0.00) (0.00)\\
\hline
\hline
   total     &                &351.18 (2.04) (9.51)\\
\hline
\hline
\ent

\newpage

We now turn to the hadronic contribution to the shift in the
fine structure constant. In Table~5a we list the contributions from
different energy ranges.
\bet{|c|c|r|r|}
\multicolumn{4}{c}{Table~5a: Contributions to 
$\dalh \times 10^4$}\\
\hline
\hline
 final state &energy range (GeV) &contribution (stat) (syst)
& \multicolumn{1}{c|}{Ref.~\ci{TASI}}\\
\hline
  $ \rho   $     &(0.28, 0.81)      & 26.08[26.23] (0.29) (0.62)
&26.07(0.10)(0.78)\\
  $ \omega $     &(0.42, 0.81)      &  2.93[ 2.96] (0.04) (0.08)
& 3.43(0.35)(0.10)\\
  $ \phi   $     &(1.00, 1.04)      &  5.08[ 5.15] (0.07) (0.12)
& 5.27(0.24)(0.16)\\
  $ J/\psi $     &                  & 11.34[11.93] (0.55) (0.61)
&10.16(1.34)(1.52)\\
  $\Upsilon$     &                  &  1.18[ 1.27] (0.05) (0.06)
& 1.17(0.04)(0.07)\\
  hadrons        &(0.81, 1.40)      & 13.83[13.99] (0.15) (0.79)
&15.63(0.68)(1.73)\\
  hadrons        &(1.40, 3.10)      & 27.62[28.23] (0.32) (4.01)
&27.95(0.60)(5.59)\\
  hadrons        &(3.10, 3.60)      &  5.82[ 5.98] (0.30) (1.12)
& 5.98(0.31)(1.15)\\
  hadrons        &(3.60, 9.46)      & 50.60[50.50] (0.24) (3.33)
&50.27(0.51)(3.14)\\
  hadrons        &(9.46, 40.0)      & 93.07~~~~~~~~ (0.86) (3.39)
&93.60(1.24)(2.91)\\
  perturbative   &(40.0, $\infty$ )   & 42.82~~~~~~~~ (0.00) (0.10)
&42.67(0.29)(0.59)\\
\hline
\hline
   total       &                    &280.37[282.13](1.18) (6.43)
&282.21(2.19)(8.30)\\
\hline
\hline
\ent
\bet{c|c|c|c|c|cc}
\multicolumn{6}{c}{Table~5b: Shape dependence of resonance
contributions}\\
\hline
 final state & energy range (GeV) &  a)    &  b)    &  c)    &  d)   & \\
\hline
  $ \omega $ &(0.42, 0.81)    &  3.138 &  2.994 &  2.952 &  2.931 &\\
  $ \phi $   &(1.00, 1.04)    &  5.451 &  5.068 &  5.069 &  5.083 &\\
  $ J/\psi $ &                & 11.380 & 11.342 & 11.338 & 11.338 &\\
  $\Upsilon$ &                &  1.182 &  1.178 &  1.178 &  1.178 &\\
\hline
\multicolumn{6}{l}{a) narrow width approximation}&\\
\multicolumn{6}{l}{b) non-relativistic} &\\
\multicolumn{6}{l}{c) relativistic constant width}&\\
\multicolumn{6}{l}{d) relativistic $s$-dependent width} &\\
\ent
The resonance contributions were evaluated as discussed in
Sec.~\ref{sec:RESO} Note that Breit-Wigner resonances may be treated
non-relativistically, relativistically and with different
off-resonance behavior. Results obtained for different resonance
shapes are listed in Tab.~5b.  We note that the narrow width
approximation gives generally larger values than the Breit-Wigner
parametrizations in either the non-relativistic form, the relativistic
form with constant width or the relativistic form with $s$-dependent
width. As expected, the deviations obtained from using different types
of Breit-Wigner parametrizations are within the experimental
uncertainties.

The region between the $J/\psi$ and the $\Upsilon$ has been split into
two subdomains. From the $J/\psi$ to 3.6 GeV only Mark I data are
available. Above 3.6 GeV up to the $\Upsilon$ one may think about
skipping the Mark I data, as mentioned before. Because of the
resonances included in the data we integrate the PLUTO, DASP and Mark
I data separately in the region from 3.6 to 5.2 GeV and combine the
results after integration.

The analysis shows that the results are affected in a minor way if we
include the Mark I data. Taking the weighted average of the two values
obtained from the Mark I data on the one hand and the other data on the
other hand we find 50.79(0.20)(3.20). If we combine results from individual
experiments and take weighted averages for overlapping domains we get
50.69[51.20](0.30)(2.74) if we exclude [include] the Mark I data. These
checks show that the trapezoidal rule works consistently and there is no
serious problem of properly weighting the data from different experiments
according to their uncertainties\footnote{This is in contrast to claims in
Ref.~\ci{Swartz94} that the trapezoidal rule does not allow us to combine
data from different experiments in a reasonable way and automatically has
the effect of weighting all inputs equally. Note that in previous
analysis~\ci{Jegerlehner86,Burkhardt89,TASI} properly weighted averages of
data from different experiments were used unless data from different
experiments had very similar errors in which case they were treated like
points from one experiment. As systematic errors for a given ``experiment''
are added linearly the simplified treatment of equal weighting of data has
primarily the effect of yielding a too conservative estimate for the
systematic error.  In Ref.~\ci{Swartz94} data are fitted to smooth
functions before integration and systematically lower results are obtained
(see also the discussion in Ref~\ci{pitfal}).  The author of
Ref.~\ci{Swartz94} believes more in integration of his fits than in
trapezoidal integration, without giving any actual arguments; moreover he
does not mention the fact that his fitting method has to rely on some
rather arbitrary assumptions about the sources of systematic effects and/or
correlations.}.  The values obtained compare with the ones from a previous
analysis~\ci{TASI} and are all found to be consistent.

We have checked the non-resonant part above the $J/\psi$, which
contributes a major part to $\dal$, against the 3-loop perturbative QCD
prediction, using the LEP value $\alpha_s(\mz)=0.126\pm0.005$~\ci{LEP}
for the strong interaction constant as an input. We obtain
\bea \begin{array}{cccc}
\hline
{\rm Range} & {\rm Data} 
& {\rm Ref}.~\ci{ChetyrkinKuehn95} & {\rm Eq}.~\re{Rpert}\\
\hline
 5.00-9.46 \mbox{ \ GeV} &  32.63 &  35.78 & 34.97 \\
 12.0-40.0 \mbox{ \ GeV} &  79.22 &  78.23 & 77.64 \\
\hline
\end{array}
\eea
The two QCD results differ by a different treatment of mass effects.
In Eq.~\re{Rpert} just the lowest order threshold factor is used, while
Ref.~\ci{ChetyrkinKuehn95} also takes into account charm and bottom
mass effects in the higher order terms in an expansion in $m^2/s$.

We have also varied the high energy cut energy from $E_{\rm cut}$=40
to 30, 20 and 12 GeV and found stable results: $0.0280\pm0.0007$,
$0.0280\pm0.0006$, $0.0280\pm0.0006$ and $0.0279\pm0.0006$. We note
that data and the perturbative prediction in average fit fairly well
above the resonance regions, as may be seen in Figs.~7 and 8.  In
addition we observe that the onset of the $\gamma-Z$ mixing is well
under control after the subtraction of the $\gamma-Z$ interference
term which we described at the end of Sec.~\ref{sec:dal} This
subtraction was applied already in previous
work~\ci{Burkhardt89,TASI}. Doubtless, the LEP experiments have
dramatically improved our confidence in perturbative QCD and we may
well use a much lower $E_{\rm cut}$, such as for example 12 GeV, and
use perturbative QCD also in the range from e.g. 5 GeV, up to the
$\Upsilon$ threshold. As a result we find $0.0282\pm0.0005$ and hence
a slightly larger result with a smaller error. However, if we move the
cut slightly from 5 GeV to 4.5 GeV we obtain $0.0280\pm0.0005$ which
shows that the better accuracy is delusive as the central value
depends substantially on the cut\footnote{Our findings do not confirm
those of Ref.~\ci{MZ94}.  These authors essentially use perturbative
QCD with the world average strong coupling $\alpha_s=0.118 \pm 0.007$
for $E>3$ GeV, plus the resonances, instead of the data. Data, in so
far as they are used, are rescaled to meet the result of perturbation
theory. Their results lie systematically lower than ours. The third
order QCD prediction is assumed to be exact down to 3 GeV.}. A glance
at Fig.~7a shows that the PLUTO data points in the region of the
$\psi(4040)$, $\psi(4160)$ and $\psi(4415)$ resonances are lower than
``perturbative QCD plus the Breit-Wigner resonances''.  Note that the
resonances depicted in Fig.~7a have been scaled down by
$(\al/\al(s))^2$ in order to have the proper normalization for $R$. If
one applies the missing $(1+2 \Delta
\al_l) (\alpha/\alpha(s))^2$ correction to the Mark I data they agree
much better than the PLUTO data with the above ``prediction''.

As a main result we obtain
\ba
  \dalh &=& 0.0280[0.0282] \pm 0.0007
\label{dalpres}
\ea
for $M_Z=91.1888$~GeV consistent with the 1991 update \ci{TASI}. In
brackets we also give the value without any $(\al/\al(s))^2$ rescaling
of data. These results confirm the assumption made in previous
work, namely, that corrections which account for the missing
subtractions of the vacuum polarization contributions are small and
well within errors. In fact usually the ``missing corrections'' were
estimated and included as a part of the systematic error. This small
correction was not taken into account in \ci{TASI}.

A lower central value than in~\ci{Jegerlehner86,Burkhardt89} is mainly
due to the new more precise data from CB~\ci{Edwards90} and
MD-1~\ci{Blinov91}. The Mark I data dominated in most of the
earlier estimates. The replacement of the Mark I data by the CB data
in the common energy range has been presented in the update
\ci{TASI} some time ago.  Undressing of the resonance contributions and some
supplementary subtractions of hadronic vacuum polarization
contributions lead to a correction of $-0.000176$. We now employ the value
for $\alpha_s$ from LEP. In Refs.~\ci{Jegerlehner86,Burkhardt89,TASI}
the $\alpha_s$ value used was the one obtained by the PEP/PETRA
experiments.

The uncertainty obtained in this analysis is smaller mainly for the
following reasons. First, we are using more complete data in the range
from 0.81 to 3.1 GeV. In particular the DM2 data~\ci{Schioppa86} in
the range from 1.35 to 2.3 GeV, which have smaller uncertainties, were
not used in previous estimates. Second, previously in
\ci{Burkhardt89,TASI} an overall 20\% systematic error was assumed in
this range, which corresponded to a typical systematic error reported
by the individual experiments. Furthermore, in these references, a
conservative 3.5\% error, which was the accuracy of the PEP/PETRA
data, was assumed for the perturbative tail. This more conservative
treatment of errors in Ref.~\ci{Burkhardt89} lead to an increase to
0.0009 from 0.0007 which was previously obtained in
Ref.~\ci{Jegerlehner86}. The more complete collection of data allows
us to better justify the more precise result now. Although the
replacement of the Mark I data by the CB data helps in reducing the
uncertainty it is not the main reason as can be checked in Tab.~5a.
The higher accuracy of the resonance parameters also helped in
reducing the uncertainty.

We finally summarize in Tab.~5c the uncertainties obtained for the
different contributions.
\bet{|c|r|c|c|}
\multicolumn{4}{c}{Table~5c: ``Distribution'' of uncertainties} \\
\hline
        & $\dalh \times 10^4$ &rel. err.&abs.err.\\
\hline
          Resonances:           &   46.61 (1.08) &   2.3 \% &   0.4 \%
\\
          Background:           &                &          &

\\
         $E<M_{J/\psi}$         &   41.45 (4.11) &   9.9 \% &   1.5 \%
\\
     $M_{J/\psi}<E<3.6$ GeV     &    5.82 (1.16) &  19.9 \% &   0.4 \%
\\
   $3.6$ GeV $<E<M_{\Upsilon}$  &   50.60 (3.34) &   6.6 \% &   1.2 \%
\\
  $M_{\Upsilon}<E<$   40 GeV    &   93.07 (3.50) &   3.8 \% &   1.2 \%
\\
\hline
      $E < $ 40 GeV data        &  237.55 (6.54) &   2.8 \% &   2.3 \%
\\
      40 GeV $ < E$ QCD         &   42.82 (0.10) &   0.2 \% &   0.0 \%
\\
\hline
     total                &  280.37 (6.54) &\multicolumn{2}{c|}{ 2.3 \%} 
\\
   $(\star)$              &         (4.92) &\multicolumn{2}{c|}{(1.8 \%
)} \\
\hline
\ent

The last line ($\star$) of Tab.~5c gives the uncertainty one would get
if the experimental errors on $R(s)$ would be reduced to 5\% in all
the regions which exhibit uncertainties larger than that. The last
column gives the uncertainty relative to the total result. The table
clearly tells us that the background contributions require a new scan
for all energies up to about 12 GeV. For higher energies
one may rely more on perturbative QCD. It seems unlikely that a
substantial improvement will be possible in the foreseeable future.
Physics will need a global update of many experiments in
order to be prepared for the next level of precision physics.

\section{Summary and outlook}

The question at the beginning of this investigation was whether one
could improve the previous estimates in the present situation and how
things could develop in future. While there has been very little truly
new experimental results, some groups have published updated results
which were available before as preliminary data only. Examples are the
ND results~\ci{Dolinsky91}, the DM2
results~\ci{Antonelli92,Schioppa86,Bisello91,Bisello83} and the
Crystal Ball results~\ci{Edwards90}, where the latter have already
been used in~\ci{TASI}. New results from VEPP-4~\ci{Blinov91} have
been included as well. In addition we have made an effort to collect
as much as possible all the available data. It should be noted that
for example the Durham-RAL High Energy Physics database utilized in
Ref.~\ci{MZ94} is incomplete and in many cases contains preliminary
data only while the final results are missing. In our analysis we also
used data which have not yet been included in the
collection~\ci{Landolt} of $\epm$ data which was published recently.
Another issue was to check what corrections were applied to the
published data, which required a careful reading of the original
papers.

Additional motivations for performing this update were the following:
the recent issues of the Review of Particle Properties~\ci{PDG}
had some improvements on the resonance parameters and there was
progress in calculating $R(s)$ at high energies in perturbative
QCD~\ci{Rpert3,ChetyrkinKuehn95,LEP}. For the muon anomaly the study
of the low energy end by means of chiral perturbation
theory~\ci{Juerg} allowed us to reduce potential model-dependences of
earlier approaches.

That all these efforts mentioned lead to minor changes of the results
was to be expected.  Nevertheless we think such an update was
necessary and it will not be soon that a more precise estimate will be
possible.
  
Refinements and improvements were proposed recently in
Refs.~\ci{Swartz94} and \ci{MZ94}. More theoretical attempts to
calculate the photon vacuum polarization may be found in
\ci{Chang82,Scheck84,Okun94,Geshkenbein94,Pallante94} to mention only a
few. We all know about the difficulties to make accurate {\em
reliable} predictions in strong interaction physics. Here we have
tried to estimate hadronic vacuum polarization effects in a 
model-independent way exploiting as much as possible the existing data, and
we find results consistent with known estimates.

For the muon anomaly we propose to use the RG improved value which is
 1$\sigma$ higher than the bare one usually considered.  With our
value for $a_\mu^{{\rm had}\:*}$, and assuming that the subleading
hadronic contributions and their uncertainties are as given by
Kinoshita et al.~\ci{Kino2}, we obtain $$a_\mu^{{\rm the}} = (11 659
210 \pm 16)\times 10^{-10}$$ and therefore at present we have
$$a_\mu^{{\rm exp}}-a_\mu^{{\rm the}}=(20 \pm 86)\times 10^{-10}\;.$$
The forthcoming Brookhaven experiment is expected to reduce this
uncertainty to $\pm 17\times 10^{-10}\:.$ Recently there is an ongoing
discussion about the true size of the hadronic light--by--light
scattering contribution~\ci{Barbi} and therefore the value and the
uncertainty of the theoretical prediction still may change.

Finally, by adding the hadronic contribution, the leptonic
contribution $\Delta \al _{l}=0.031421$ and a top contribution $\Delta
\al _{top}=-0.000061$ (obtained by integrating Eq.~\re{Rpert} with
running parameters numerically) we find the total shift by the
fermions\footnote{We do not include the $W$ boson contribution here
because it is gauge dependent, and splitting off a gauge invariant part
and combining the reminder with the photon vertex is not
unique~\ci{JEFL}.} $$\dal=0.05940 \pm 0.00065$$ and hence for the
effective fine structure constant $$\al (\mz)^{-1}=128.896 \pm
0.090\;\;.$$

What about the future?
     Progress in decreasing the hadronic uncertainties can be expected
from the experiments at the general purpose CMD-2 detector at the
Novosibirsk VEPP-2M collider, which has an ambitious goal of measuring
$\sigma_{tot}(e^+e^- \to hadrons)$ with an accuracy better than $1 \%$.
The experiment is in progress and during the 1995 runs it is planned
to study the energy range from the threshold of two pion production up
to the $\phi$ meson with a 10 MeV step and achieve a statistical
accuracy of $3 \%$ in each point~\ci{Khazin94}. Further work will be
needed to understand the detector performance as well as low energy
radiative corrections at such a level that systematic uncertainties will
be understood with the desired accuracy. After that the energy range
from the $\phi$ meson up to the maximum attainable energy of 1.4 GeV
will be studied. Here, as discussed above, additional difficulties can
arise for the high precision cross section measurements because of the
intermediate mechanism uncertainty. Let us assume that a systematic
uncertainty of $1 \%$ will be achieved for the $\rho , \omega$ and
$\phi$ mesons and that of $2 \%$ for the hadron continuum between
0.81 and 1.40 GeV. The statistical error in the integrals can be
neglected and a resulting systematic uncertainty in the muon anomaly
decreases from $15\times 10^{-10}$ to $9\times 10^{-10}$ of which $8
\times 10^{-10}$ come from the contribution
of the hadron continuum between 1.4 and 3.1 GeV.
       
     Further progress can be expected from the future generation
experiments at DA$\Phi$NE at Frascati and the upgraded VEPP-2M.  Plans
at DA$\Phi$NE are to perform a scan with 100 values from 0.28 to 1.5
GeV c.m. energy with a precision which will allow to reduce the
uncertainty from the hadronic vacuum polarization contribution to
$a_\mu$ to about 0.3\%~\ci{Franzini95}. This will reduce the
uncertainty for the region below 1.4 GeV to $1.5\times 10^{-10} $ and
the remaining total hadronic vacuum polarization uncertainty in the
prediction of $a_\mu$ would be $8\times 10^{-10} $, which again is
completely dominated by the contribution from 1.4 to 3.1 GeV.  The
DA$\Phi$NE measurement will be possible earliest by end of 1997 but
then can be done within a few days. Obviously, the VEPP-2M and the
DA$\Phi$NE measurements will be crucial for the physics we will be able
to learn from the planned Brookhaven Experiment 821~\ci{BNL}.

No improvement is in sight at higher energies. The energy region from
1.4 to 3.1 GeV and in addition that of higher energies from 3.1 up to
about 12 to 40 GeV, depending on how much one is accepting to rely on
perturbative QCD, will give a dominant contribution to the uncertainty
of the fine structure constant shift.

     One can conclude that a real breakthrough in improving the precision
of the hadronic vacuum polarization will require dedicated efforts in
high precision $R$-measurements in a wide energy range.   

\bigskip

{\bf Acknowledgments}

One of the authors (S.E) would like to thank the Paul Scherrer Institute
where part of this work was done for its hospitality, and 
Boris Khazin, Georgi Shestakov and Boris Shwartz for useful discussions.
We also thank Rinaldo Baldini for providing us sets of $R$--data and Paula
Franzini for helpful discussions and for carefully reading the manuscript.

\bigskip

\large

{\bf Appendix: ``Dressed'' versus ``undressed'' quantities}

\normalsize

Equations~\re{amu} and~\re{dal} can be derived from the convergent dispersion
relation
\bea
{\rm Re}\: \Delta \hat{\Pi}'_{\gamma}(s)=
{\rm Re}\: \hat{\Pi}'_{\gamma}(s)-\hat{\Pi}'_{\gamma}(0)=
\frac{s}{\pi} {\rm \ Re} \int_{s_0}^{\infty}
ds' \frac{{\rm \ Im}\: \hat{\Pi}'_{\gamma}(s')}{s'(s'-s -i \veps)} \nn
\eea
where $\hat{\Pi}'_\g$ is the transversal part of the current-current
matrix element of the conserved electromagnetic current
\bea
\hat{\Pi}_{\mu \nu}(q)=i\: \int d^4x\: e^{iqx}\:
<0|\: T^*\: j^\gamma_\mu(x)\: j^\gamma_\nu(0)\:|0>
=(q^2 g_{\mu \nu}-q_\mu q_\nu) \hat{\Pi}'_\g (q^2)\;\;.
\eea
This self-energy function exists in the limit where the electromagnetic
interaction is switched off and for the hadronic current, in
principle, it is determined by QCD.  The {\em irreducible} photon
self-energy is given by $\Pi'_\g (s)=e^2 \hat{\Pi}'_\g (s)$. It is
well known that the running of $\alpha=e^2/4\pi$ (see Sec.~3) may be
understood as a consequence of the Dyson summation of the irreducible
photon self-energy
\bea
\frac{e^2}{s}\:\left( 1+{\rm Re}\: \Delta \Pi'_\g (s)\right) \ra
\frac{e^2}{s}\:{\rm Re}\: \frac{1}{1-\Delta \Pi'_\g (s)} \equiv
\frac{e^2(s)}{s}
\eea
which is equivalent to a RG improvement of the perturbation expansion.
For the full photon propagator we have
\bea
{\rm Re}\: \frac{1}{1-\Delta \Pi'_\g (s)}=
\frac{1}{1-{\rm Re}\: \Delta \Pi'_\g (s)}
\frac{1}{1+(\frac{{\rm Im}\: \Pi'_\g (s)}{1-{\rm Re}\: \Delta \Pi'_\g (s)})^2}
\simeq
\frac{1}{1-{\rm Re}\: \Delta \Pi'_\g (s)} = \frac{e^2(s)}{e^2}
\eea
for the real part and
\bea
{\rm Im}\: \frac{1}{1-\Delta \Pi'_\g (s)}=
\frac{{\rm Im} \Pi'_\g (s)}{(1-{\rm Re}\: \Delta \Pi'_\g (s))^2}
\frac{1}{1+(\frac{{\rm Im}\: \Pi'_\g (s)}{1-{\rm Re}\: \Delta \Pi'_\g (s)})^2}
\simeq
{\rm Im}\: \Pi'_\g (s) \frac{e^4(s)}{e^4}
\eea
for the imaginary part.

These relations show that in the dispersion relation for the irreducible
self-energy one has to use ``undressed'' (with respect to vacuum
polarization effects) quantities and not the physical ones. Thus in
the optical theorem (unitarity) one has to use the undressed total
cross section
\bea
-{\rm Im}\: \hat{\Pi}'_{\gamma}(s)
=\frac{s}{e^4(s)} \sigma_{had} (s)
=\frac{s}{e^4} \sigma^{(0)}_{had} (s)
=\frac{R(s)}{12 \pi}\;\;.
\eea
Up to subleading higher order effects the resummed photon propagator
may be obtained directly from a dispersion relation by using the
dressed physical quantities under the dispersion integral.

\newpage

\bb{99}

\bi{amuexp}
 J. Bailey et al. \pl{68} (1977) 191;\\ F.J.M. Farley and E. Picasso,
 Ann. Rev. Nucl. Sci. {\bf 29} (1979) 243; in ``Quantum Electrodynamics'',
 ed. T. Kinoshita, World Scientific, Singapore, F.J.M. Farley, 
 \zp{56} (1992) S88.

\bi{Kino1}
 T. Kinoshita, {\em Past, Present and Future of Lepton $g-2$},
 in Proc. of the 10th Inter. Symposium on ``High Energy Spin Physics'',
 Nagoya, Japan, 1992; \zp{56} (1992) S80;\\
 T. Kinoshita, in ``Quantum Electrodynamics'',
 ed. T. Kinoshita, World Scientific, Singapore, 1990, pp. 218 - 321.

\bi{KiLi}
 T. Kinoshita and W.B. Lindquist, \prd{41} (1990) 593;\\
 M.A. Samuel and G. Li, \prd{44} (1991) 3935;\\
 S. Laporta and E. Remiddi, \plb{301} (1993) 440.

\bi{BKT}
 D.J. Broadhurst, A.L. Kataev and O.V. Tarasov, \plb{298} (1993) 445;\\
 T. Kinoshita, \prd{47} (1993) 5013.

\bi{KiMa}
 T. Kinoshita and W. J. Marciano, in ``Quantum Electrodynamics'',
 ed. T. Kinoshita, World Scientific, Singapore, 1990, pp. 419 - 478,
 and references therein.

\bi{Mars}
 P. Mery, S.E. Moubarik, M. Perrottet and F.M. Renard,
 \zp{46} (1990) 229.

\bi{Bernr}
 W. Bernreuther, \zp{56} (1992) S97.

\bi{AEW}
 C. Arzt, M.B. Einhorn and J. Wudka, \prd{49} (1994) 1370.

\bi{tau1}
 M.A. Samuel, G. Li and R. Mendel, \prl{67} (1991) 668.

\bi{tau2} 
 M. Benmerrouche, G. Orlandini and T.G. Steele, \plb{316}
 (1993) 381.

\bi{L3} 
 L. Vuilleumier, Ph.D. Thesis, University of Lausanne, 1994.

\bi{BNL}
 B. Lee Roberts (BNL E821), \zp{56} (1992) S101.

\bi{BM}
 C. Bouchiat and L. Michel, {\em J. Phys. Radium} {\bf 22} (1961) 121.

\bi{Durand}
 L. Durand, III., \pr{128} (1962) 441.

\bi{GdeR}
 M. Gourdin and E. de Rafael, \np{10} (1969) 667.

\bi{Barger75}
 V. Barger, W.F. Long and M.G. Olsson, \pl{60} (1975) 89.

\bi{Barkov85}
 L.M. Barkov et al. (OLYA, CMD), \npb{256} (1985) 365.

\bi{Kino2}
 T. Kinoshita,  B.  Ni\v{z}i\'{c} and Y. Okamata,  \prl{52}  (1984) 717;\\
 \prd{31} (1985) 2108.

\bi{CLY}
 J.A. Casas, C. L\'{o}pez and F.J. Yndur\'{a}in,  \prd{32} (1985) 736.

\bi{MD}
 \v{L}. Martinovi\v{c} and S. Dubni\v{c}ka, \prd{42} (1990) 884.

\bi{jegup} 
 F. Jegerlehner, private communication to V.W. Hughes, 1991.

\bi{DD}
 A.Z. Dubni\v{c}kov\'{a}, S. Dubni\v{c}ka and P. Strizenec,
 {\em New Evaluation of Hadronic Contributions to the Anomalous Magnetic
 Moment of Charged Leptons}, Dubna-Report, JINR-E2-92-281 (Dec. 1992).

\bi{LEP}
 D. Schaile (LEP Collaboration), CERN-PPE-94-162, 1994; Talk given at
 27th International Conference on High Energy Physics (ICHEP), Glasgow,
 Scotland, 20-27 July 1994.

\bi{Jegerlehner86}
 F. Jegerlehner, \zp{32} (1986) 195.

\bi{LynnPensoVerz}
 B. W. Lynn, G. Penso, C. Verzegnassi, \prd{35} (1987) 42.

\bi{Burkhardt89}
 H. Burkhardt, F. Jegerlehner, G. Penso, C. Verzegnassi, \zp{42} (1989) 497.

\bi{TASI} 
 F. Jegerlehner, in ``Testing the Standard Model'', \\  eds. M.
 Cveti\v{c}, P. Langacker, World Scientific, Singapore, 1991, p. 476;
 {\em Prog. Part. Nucl. Phys.} {\bf 27} (1991) 32.

\bi{Swartz94}
 M.L. Swartz, Preprint SLAC-PUB-6710, 1994.

\bi{MZ94}
 A.D. Martin and D. Zeppenfeld, Preprint MAD/PH/855, 1994.

\bi{Berends76}
 F.A. Berends and G.J. Komen, \pl{63} (1976) 432;\\
 E.A. Paschos, \npb{159} (1979) 285;\\
 J. Ellis, M.K. Gaillard, D.V. Nanopoulos and S. Rugaz, \npb{176} (1980) 61;\\
 W. Wetzel, \zp{11} (1981) 117.

\bi{Barbi}
 R. Barbieri and E. Remiddi, in ``The DA$\Phi$NE physics handbook''
 Vol. II, \\ eds. L. Maiani et al., (INFN, Frascati, 1992) p. 301;\\
 M.B. Einhorn, \prd{49} (1994) 1668;\\
 E. de Rafael, \plb{322} (1994) 239.

\bi{Rpert3}
 S.G. Gorishny, A.L. Kataev and S.A. Larin, \plb{259} (1991) 144;\\
 L.R. Surguladze, M.A. Samuel, \prl{66} (1991) 560; idid., 2416(E).

\bi{ChetyrkinKuehn95}
 K.G. Chetyrkin and J.H. K\"uhn, \plb{342} (1995) 356 and references therein.

\bi{BonneauMartin}
 G. Bonneau and F. Martin, \npb{27} (1971) 381;\\
 D.R. Yennie, \prl{34} (1975) 239;\\
 J.D. Jackson and D.L. Scharre, {\em Nucl. Instrum. Methods} {\bf 128} 
 (1975) 13;\\
 M. Greco, G. Pancheri-Srivastava and Y. Srivastava, \npb{101} (1975) 234;\\
 {\bf B202} (1980) 118.

\bi{Tsai}
 Y.S. Tsai, SLAC-PUB-1515 (1975); SLAC-PUB-3129 (1983).

\bi{BerendsKleiss}
 F.A. Berends et al., \npb{57} (1973) 381; \npb{68} (1974) 541;\\
 F.A. Berends and R. Kleiss, \npb{178} (1981) 141;\\
 F.A. Berends, R. Kleiss and S. Jadach, \npb{202} (1982) 63;\\
 F.A. Berends and R. Kleiss, \npb{228} (1983) 537.

\bi{Eidelman78}
 S.I. Eidelman and E.A. Kuraev, \pl{80} (1978) 94;\\
 V.N. Baier, V.S. Fadin, V.A. Khoze and E.A. Kuraev, \PR{78} (1981) 293;\\
 E.A. Kuraev and V.S. Fadin, \SJNP{41} (1985) 466.

\bi{BerBoe}
 F.A. Berends and A. B\"ohm,  in ``High Energy
 Electron-Positron Physics'',\\ eds.  A. Ali and P. S\"oding, World
 Scientific, Singapore, 1988, p. 27-140 (see Fig.~8.5).

\bi{CabbiboGatto61}
 N. Cabbibo and R. Gatto, \prl{4} (1960) 313, \pr{124} (1961) 1577.

\bi{gusa}
 G. Gounaris, J. Sakurai, \prl{21} (1968) 244;\\
 M. Gourdin et al., \pl{30} (1969) 347;\\
 A. Quenzer et al., \pl{76} (1978) 512.

\bi{PDG}
 L. Montanet et al. (Review of Particle Properties), \prd{50} (1994) p.1, 1173.

\bi{PDG88}
 G.P. Yost et al. (Review of Particle Properties), \plb{204} (1988) p.339.

\bi{Buchmueller88}
 W. Buchm\"uller and S. Cooper, in ``High Energy Electron-Positron
 Physics'', \\ eds.  A. Ali and P. S\"oding, World Scientific, Singapore,
 1988, p. 410-487.

\bi{Koenigsmann87}
 K. K\"onigsmann, DESY 87-046 (1987).

\bi{Alexander88}
 J.P. Alexander, G. Bonvicini, P.S. Drell and R. Frey, \prd{37} (1988) 56;\\
 J.P. Alexander, G. Bonvicini, P.S. Drell, R. Frey and V. L\"uth,
 \npb{320} (1989) 45.

\bi{Aulchenko87}
 V.M. Aulchenko et al., \pl{186} (1987) 432.

\bi{Achasov76}
 F.M. Renard, \npb{82} (1974) 1;\\
 N.N. Achasov, N.M. Budnev, A.A. Kozhevnikov and G.N. Shestakov,
 \SJNP{23} (1976) 320;\\
 N.N. Achasov, A.A. Kozhevnikov, M. S. Dubrovin, V.N. Ivanchenko and
 E. V. Pakhtusova, \IJMP{7} (1992) 3187.

\bi{Juerg}
 J. Gasser, U.-G. Mei\ss ner, \npb{357} (1991) 90.

\bi{Amendolia86}
 S.R. Amendolia et al. (NA7), \npb{277} (1986) 168.

\bi{Augustin79}
 J.E.Augustin, Proc. EPS Int.Conf.on HEP, CERN, Geneva, 1979;\\
 Proc. of the 1979 Carg\'ese Summer Institute on Quarks and Leptons,\\
 eds. M. L\'evy et al., Plenum Press, New York, 1980.

\bi{pitfal}
 G. D' Agostini, Preprint DESY 93-175, 1993.

\bi{LUND}
 T. Sj\"ostrand and M. Bengtson, Comp. Phys. Comm. {\bf 27} (1982) 243;\\
 T. Sj\"ostrand, Comp. Phys. Comm. {\bf 28} (1983) 229.

\bi{Clegg94}
 A.B. Clegg and A.Donnachie, \zp{62} (1994) 455.

\bi{Vasserman79}
 I.B. Vasserman et al. (OLYA), \SJNP{30} (1979) 519.

\bi{Vasserman81}
 I.B. Vasserman et al. (TOF), \SJNP{33} (1981) 709.

\bi{Amendolia84}
 S.R. Amendolia et al. (NA7), \pl{138} (1984) 454.

\bi{Barkov79}
 L.M. Barkov et al. (VEPP-2M), Preprint INP 79-117, Novosibirsk, 1979.

\bi{Quenzer78}
 A. Quenzer et al. (DM1), \pl{76} (1978) 512.

\bi{Bisello89}
 D. Bisello et al. (DM2), \plb{220} (1989) 321.

\bi{Barbiellini73}
 G. Barbiellini et al. ($\mu \pi$), \lnc{6} (1973) 557.

\bi{Bollini75}
 D. Bollini et al. (BCF), \lnc{14} (1975) 418.

\bi{Esposito80}
  B. Esposito et al. (MEA), \pl{67} (1977) 239;\\
 \lnc{28} (1980) 337.

\bi{Barkov89}
 L.M. Barkov et al. (CMD), Preprint INP 89-15, Novosibirsk, 1989.

\bi{Dolinsky91}
 S.I. Dolinsky et al. (ND), \PR{202} (1991) 99.

\bi{Parrour76}
 G. Cosme et al. (M2N), \pl{63} (1976) 349;\\
 G. Parrour et al. (M2N), \pl{63} (1976) 357.

\bi{Cosme79}
 G. Cosme et al. (M3N), \npb{152} (1979) 215;\\
 C. Paulot (M3N), Preprint LAL-79-14, Orsay, 1979.

\bi{Espos80}
 B. Esposito et al. (MEA), \lnc{28} (1980) 195.

\bi{Cordier80}
 A. Cordier et al. (DM1), \npb{172} (1980) 13.

\bi{Antonelli92}
 A. Antonelli et al. (DM2), \zp{56} (1992) 15.

\bi{Kurdadze86}
 L.M. Kurdadze et al. (OLYA), \JL{43} (1986) 643.

\bi{Kurdadze88}
 L.M. Kurdadze et al. (OLYA), \JL{47} (1988) 512.

\bi{Barkov88}
 L.M. Barkov et al. (CMD), \SJNP{47} (1988) 248.

\bi{Cordier82}
 A. Cordier et al. (DM1), \pl{109} (1982) 129.

\bi{Schioppa86}
 D. Bisello et al. (DM2), Preprint  LAL-90-71, Orsay, 1990;\\
 M. Schioppa, Thesis, Rome, 1986.

\bi{Bisello91}
 D. Bisello et al. (DM2), Preprint LAL-91-64, Orsay, 1991.

\bi{Cordier81}
 A. Cordier et al. (DM1), \pl{106} (1981) 155.

\bi{Pais60}
 A. Pais, Ann. Phys. N.Y. {\bf 9} (1960) 548.

\bi{Bacci79}
 C. Bacci et al. ($\gamma \gamma$2), \pl{86} (1979) 234.

\bi{Bisello81}
 D. Bisello et al. (DM1), \pl{107} (1981) 145.

\bi{Antonelli88}
 A. Antonelli et al. (DM2), \plb{212} (1988) 133.

\bi{Dolinsky86}
 S.I. Dolinsky et al. (ND), \plb{174} (1986) 453.

\bi{Ivanov81}
 P.M. Ivanov et al. (OLYA), \pl{107} (1981) 297.

\bi{Ivanov82}
 P.M. Ivanov et al. (OLYA), \JL{36} (1982) 112.

\bi{Anikin83}
 G.V. Anikin et al. (CMD), Preprint INP 83-85, Novosibirsk, 1983;\\
 E.P. Solodov, Thesis, Novosibirsk, 1984.  

\bi{Mane81}
 G. Grosdidier et al. (DM1), Preprint LAL-80-35, Orsay, 1980;\\
 B. Delcourt et al. (DM1), \pl{99}  (1981) 257;\\
 F. Mane et al. (DM1), \pl{99} (1981) 261.

\bi{Bernardini73}
 M. Bernardini et al. (BCF), \pl{46} (1973) 261.

\bi{Bisello88}
 D. Bisello et al. (DM2), \zp{39} (1988) 13.

\bi{KKpi1}
 A. Cordier et al. (DM1), \pl{110} (1982) 335;\\
 F. Mane et al. (DM1), \pl{112} (1982) 178.

\bi{Delcourt79}
 G. Bassompierre et al., \pl{68} (1977) 477;\\
 B. Delcourt et al. (DM1), \pl{86} (1979) 395.

\bi{Bisello83}
 D. Bisello et al. (DM2), \npb{224} (1983) 379; \zp{48} (1990) 23.

\bi{Antonelli94}
 A. Antonelli et al. (FENICE), \plb{334} (1994) 431.

\bi{Ambrosio80}
 M. Ambrosio et al. (B$\bar{{\rm B}}$), \pl{91} (1980) 155.

\bi{Esposito81}
 B. Esposito et al. (MEA), \lnc{19} (1977) 21; 30 (1981) 65.

\bi{Bernardini74}
 M. Bernardini et al. (BCF), \pl{51} (1974) 200; 53{\bf B} (1974) 384.

\bi{Bacino78}
 P.A. Rapidis et al. (Mark I/Lead Glass Wall), \prl{39} (1977) 526;\\
 W. Bacino et al. (Mark I/Lead Glass Wall), \prl{40} (1978) 671.

\bi{Siegrist82}
 J.L. Siegrist et al. (Mark I), \prd{26} (1982) 969;\\
 J.L. Siegrist, SLAC-Report No. 225, 1979.

\bi{DASP}
 R. Brandelik et al. (DASP), \pl{76} (1978) 361;\\
 H. Albrecht et al. (DASP), \pl{116} (1982) 383.

\bi{Criegee82}
 J. Burmeister et al. (PLUTO), \pl{66} (1977) 395;\\
 Ch. Berger et al. (PLUTO), \pl{81} (1979) 410;\\
 L. Criegee and G. Knies (PLUTO), \PR{83} (1982) 151.

\bi{Niczyporuk82}
 B. Niczyporuk, et al. (LENA), \zp{15} (1982) 299.

\bi{Edwards90}
 Z. Jakubowsky et al. (Crystal Ball), \zp{40} (1988) 49;\\
\hspace*{1mm} C. Edwards et al. (Crystal Ball), SLAC-PUB-5160, 1990.

\bi{Blinov91}
 A.E. Blinov et al. (MD-1), \zp{49} (1991) 239;\\
\hspace*{1mm} A.E. Blinov et al. (MD-1), Preprint BudkerINP 93-54, 
 Novosibirsk, 1993.

\bi{CELLO}
 H. J. Behrend et al. (CELLO), \plb{183} (1987) 400.

\bi{JADE}
 W. Bartel et al. (JADE), \pl{129} (1983) 145; 160{\bf B} (1985) 337;\\
\hspace*{1mm} B. Naroska et al. (JADE), \PR{148} (1987) 67.

\bi{MARKJ}
 B. Adeva et al. (MARK J), \prl{50} (1983) 799, 2051;\\
\hspace*{1mm} \PR{109} (1984) 131; \prd{34} (1986) 681.

\bi{TASSO}
 R. Brandelik et al. (TASSO), \pl{113} (1982) 499;\\
\hspace*{1mm} M. Althoff et al. (TASSO), \pl{138} (1984) 441 .

\bi{DHHM}
 P. Bock et al. (DHHM), \zp{6} (1980) 125.

\bi{CLEO}
 R. Giles et al. (CLEO), \prd{29} (1984) 1285;\\
\hspace*{1mm} D. Besson et al. (CLEO), \prl{54} (1985) 381.

\bi{CUSB}
 E. Rice et al. (CUSB), \prl{48} (1982) 906.

\bi{HRS}
 D. Bender et al. (HRS), \prd{31} (1985) 1.

\bi{MAC}
 E. Fernandez et al. (MAC), \prd{31} (1985) 1537.

\bi{Landolt}
Numerical Data and Functional Relationships in Science and Technology:
Group. 1:\\ {\em Nuclear and Particle Physics}.  Vol. 14: {\em
Electron - Positron Interactions}, by H. Schopper, (ed.), D.R.O.
Morrison, V.V. Ezhela, Yu.G. Stroganov, O.P. Yushchenko, V.  Flaminio,
(ed.) , M.R.  Whalley, (ed.), Springer, Berlin, 1992
(Landolt-Boernstein. New Series, 1.14).

\bi{Chang82}
 T.H. Chang, K.J.F. Gaemers and W.L.V. van Neerven, \np{202} (1982)
 407.

\bi{Scheck84}
 N.A. Papadopulos, J.A. Pe\~{n}arrocha, F.Scheck and K. Schilcher,
 \pl{140} (1984) \\ \hspace*{1mm} 213; \np{258} (1985) 1.

\bi{Okun94}
 V.A. Novikov, L.B. Okun and M.I. Vysotskii, \plb{324} (1994) 89.

\bi{Geshkenbein94}
 B.V. Geshkenbein and V.L. Morgunov, \plb{340} (1994) 185;\\
\hspace*{1mm} Preprint ITEP-70-94, 1994.

\bi{Pallante94}
 E. Pallante, \plb{341} (1994) 221.

\bi{JEFL}
 F. Jegerlehner and J. Fleischer, \pl{151} (1985) 65; 
 {\em Acta Phys. Polonica} {\bf B17} (1986) 709.

\bi{Khazin94} 
 B.I. Khazin, Talk given at 27th International Conference on High
 Energy Physics \\ \hspace*{1mm} (ICHEP), Glasgow, Scotland, 20-27 July 1994.

\bi{Franzini95}
 Paolo Franzini, {\em The Muon Gyromagnetic Ratio and $R_H$ at} DA$\Phi$NE, in
 the ``Second \\ \hspace*{1mm} DA$\Phi$NE Physics Handbook'',
 ed. L. Maiani, L. Pancheri and N. Paver, to appear, \\
\hspace*{1mm} 1995.

\eb

\newpage

\begin{figure}[thb]
\vspace*{6.5cm}
\epsffile[-100 0 75 50]{data1.pos}
\vspace*{7.4cm}
\epsffile[-100 0 75 50]{data2.pos}
\end{figure}
\begin{center}
\begin{minipage}[h]{12.3cm} \baselineskip 10truept
\begin{tabular}{ll}
{\bf Fig.~3:} &
\begin{tabular}[t]{p{118mm}}
The pion form factor $|F_\pi(s)|^2$ for 0.32~to~0.81~GeV. In this and
the following figures only the statistical error bars are shown.
\end{tabular} \end{tabular}
\end{minipage}
\end{center}

\newpage

\begin{figure}[thb]
\vspace*{6.5cm}
\epsffile[-100 0 75 50]{data3.pos}
\end{figure}
\begin{center}
\begin{minipage}[h]{12.3cm} \baselineskip 10truept
\begin{tabular}{ll}
{\bf Fig.~4:} &
\begin{tabular}[t]{p{118mm}}
Averaged value of $R$ for 0.81~to~1.4~GeV calculated from the
exclusive channels using all available data. For comparison the
$R$ values compiled in Ref.~\ci{Dolinsky91} are shown.
\end{tabular} \end{tabular}
\end{minipage}
\end{center}

\vspace*{6mm}

\begin{figure}[thb]
\vspace*{6.5cm}
\epsffile[-100 0 75 50]{data4.pos}
\end{figure}
\begin{center}
\begin{minipage}[h]{12.3cm} \baselineskip 10truept
\begin{tabular}{ll}
{\bf Fig.~5:} &
\begin{tabular}[t]{p{118mm}}
$R(s)$ in the range 1.4~to~2.3~GeV calculated from the available $R(n>2)$
compilations with the two-body channels added up.
\end{tabular} \end{tabular}
\end{minipage}
\end{center}

\newpage

\begin{figure}[thb]
\vspace*{6.5cm}
\epsffile[-100 0 75 50]{data5.pos}
\end{figure}
\begin{center}
\begin{minipage}[h]{12.3cm} \baselineskip 10truept
\begin{tabular}{ll}
{\bf Fig.~6:} &
\begin{tabular}[t]{p{118mm}}
The $R$ data from the various experiments in the energy range
3.1~to~9.6~GeV.
\end{tabular} \end{tabular}
\end{minipage}
\end{center}

\vspace*{6mm}

\begin{figure}[thb]
\vspace*{6.5cm}
\epsffile[-100 0 75 50]{data8.pos}
\end{figure}
\begin{center}
\begin{minipage}[h]{12.3cm} \baselineskip 10truept
\begin{tabular}{ll}
{\bf Fig.~8:} &
\begin{tabular}[t]{p{118mm}}
The $R$ data from the various experiments in the energy range
9.6~to~40~GeV. The two perturbation theory results 
are the $O(\al_s^3)$ predictions Eq.~\re{Rpert} and
Ref.~\ci{ChetyrkinKuehn95} 
with the LEP value $\al_s(M_Z^2)=0.126 \pm 0.005$ as input.
\end{tabular} \end{tabular}
\end{minipage}
\end{center}

\newpage

\begin{figure}[thb]
\vspace*{6.5cm}
\epsffile[-100 0 75 50]{data6.pos}
\vspace*{7.4cm}
\epsffile[-100 0 75 50]{data7.pos}
\end{figure}
\begin{center}
\begin{minipage}[h]{12.3cm} \baselineskip 10truept
\begin{tabular}{ll}
{\bf Fig.~7:} &
\begin{tabular}[t]{p{118mm}}
The $R$ data from the various experiments in the energy range
3.1~to~9.6~GeV.  The ``weighted mean'' does not include the Mark I
data above 3.6 GeV. The two perturbation theory results 
are the $O(\al_s^3)$ predictions Eq.~\re{Rpert} and
Ref.~\ci{ChetyrkinKuehn95} 
with the LEP value $\al_s(M_Z^2)=0.126 \pm 0.005$ as input.
\end{tabular} \end{tabular}
\end{minipage}
\end{center}

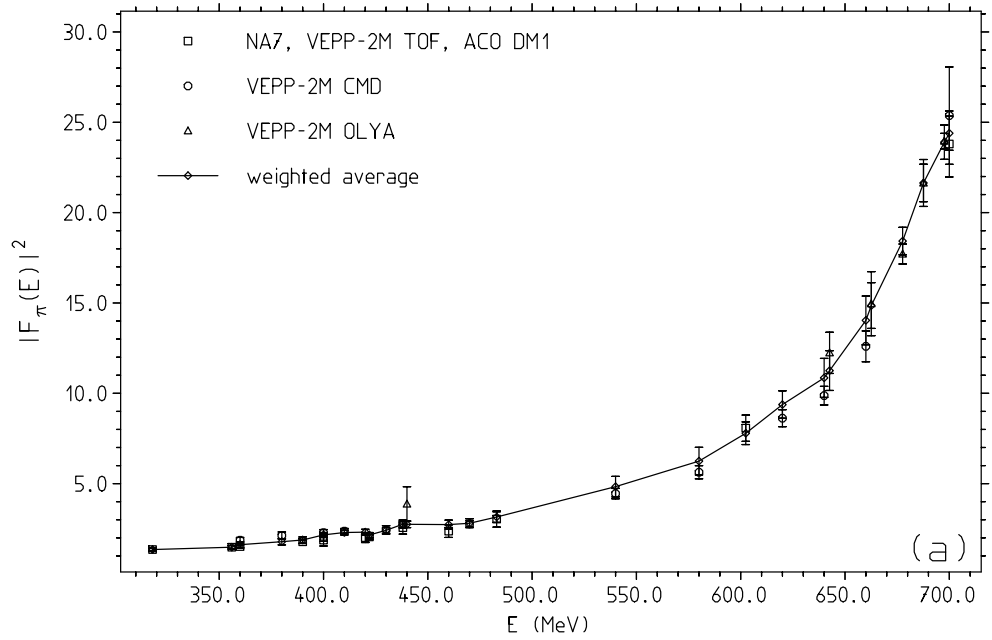
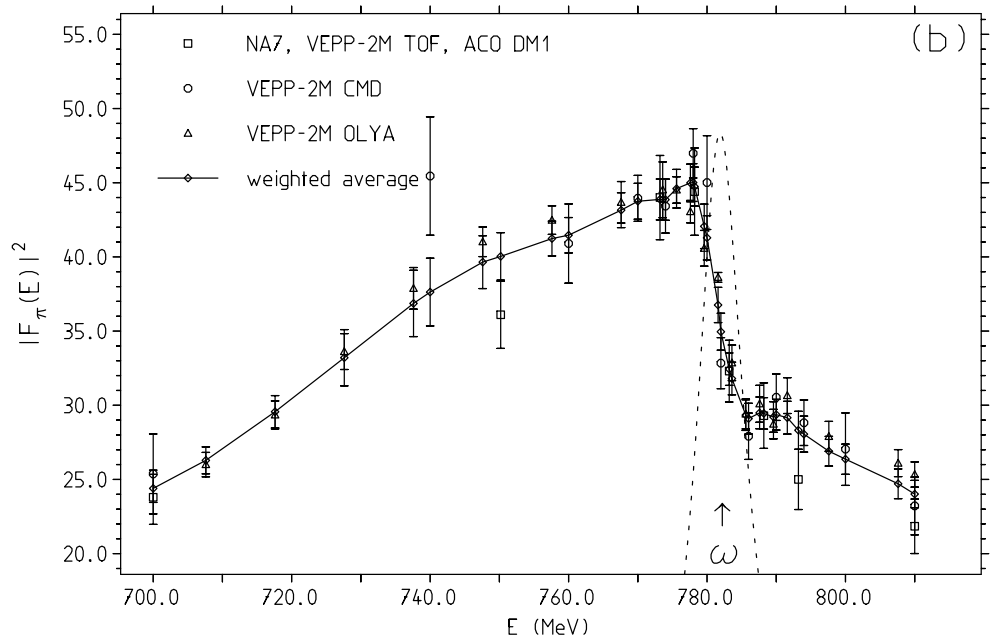
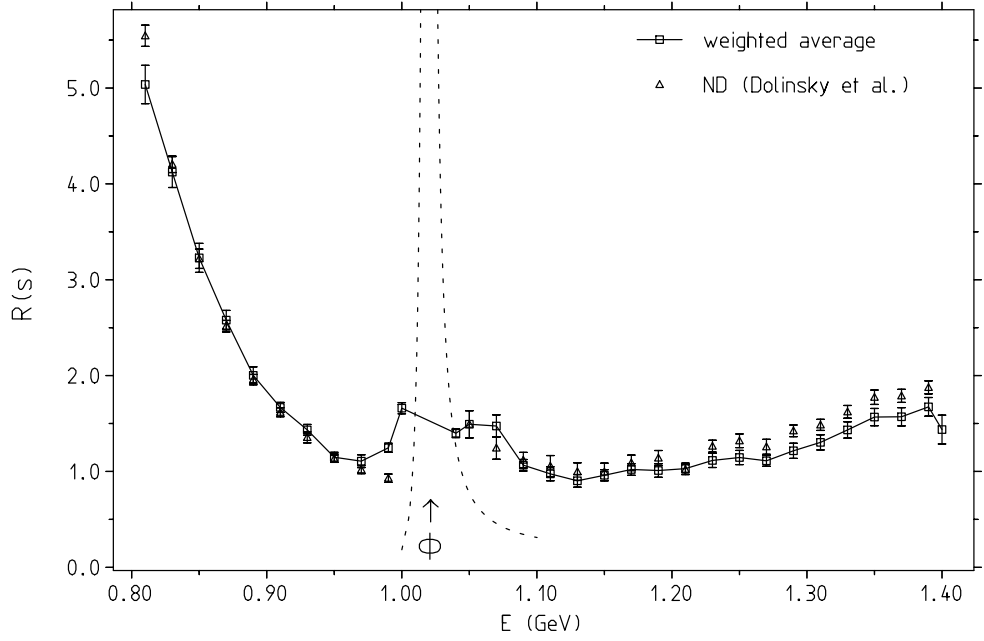
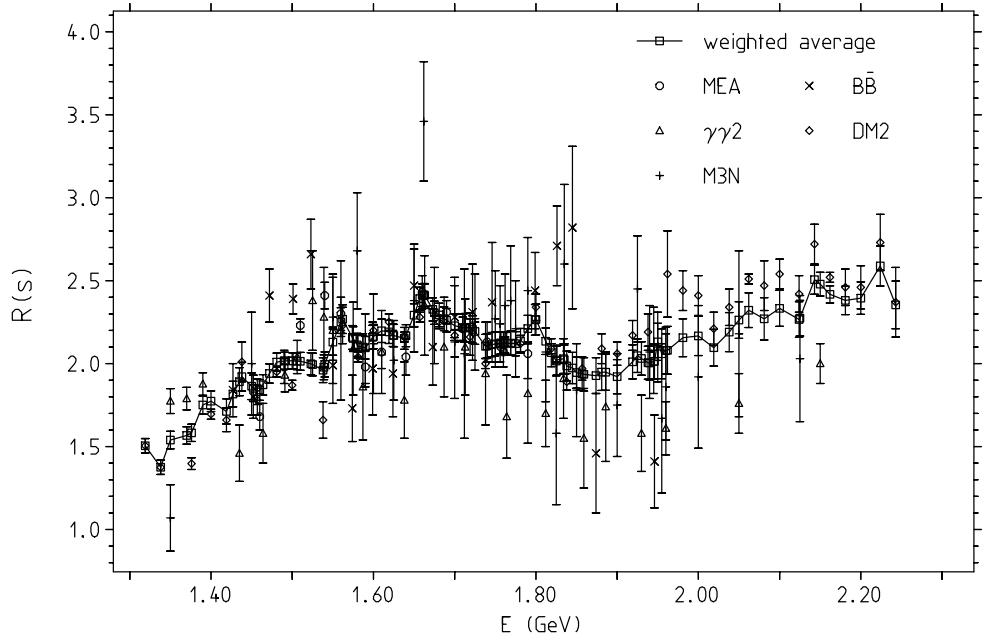
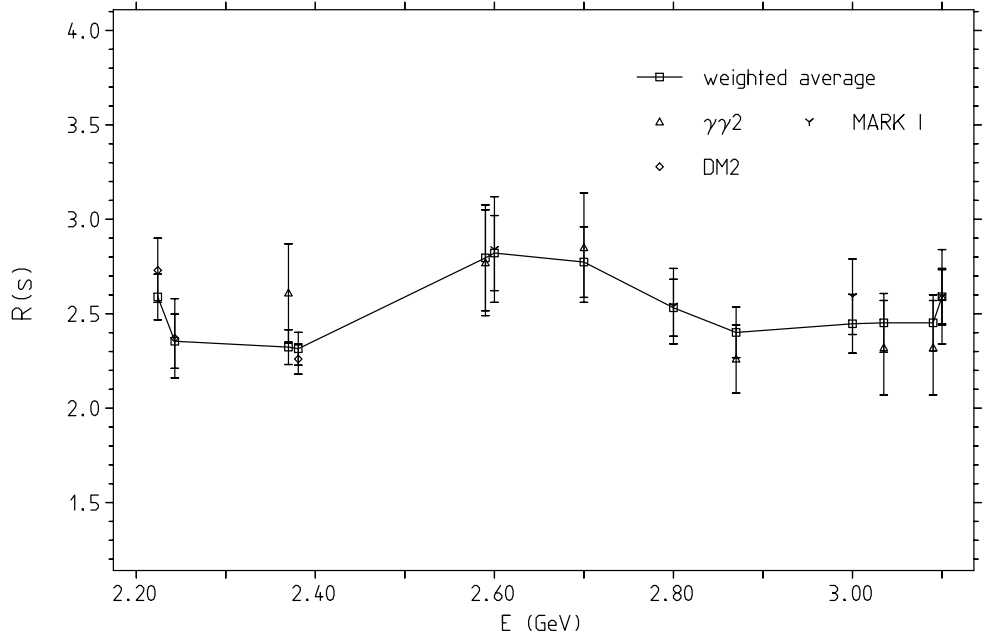
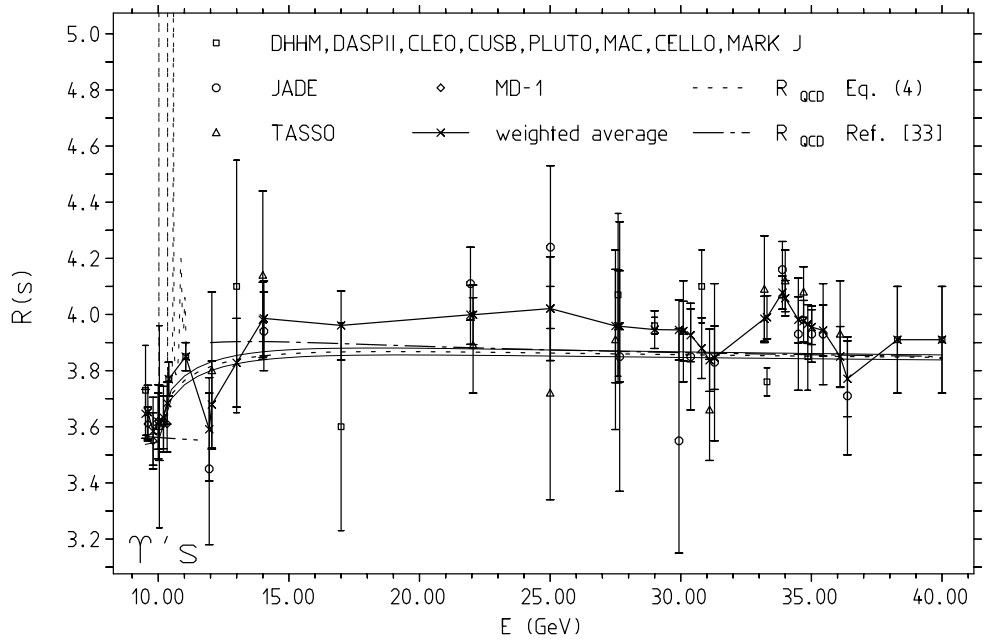
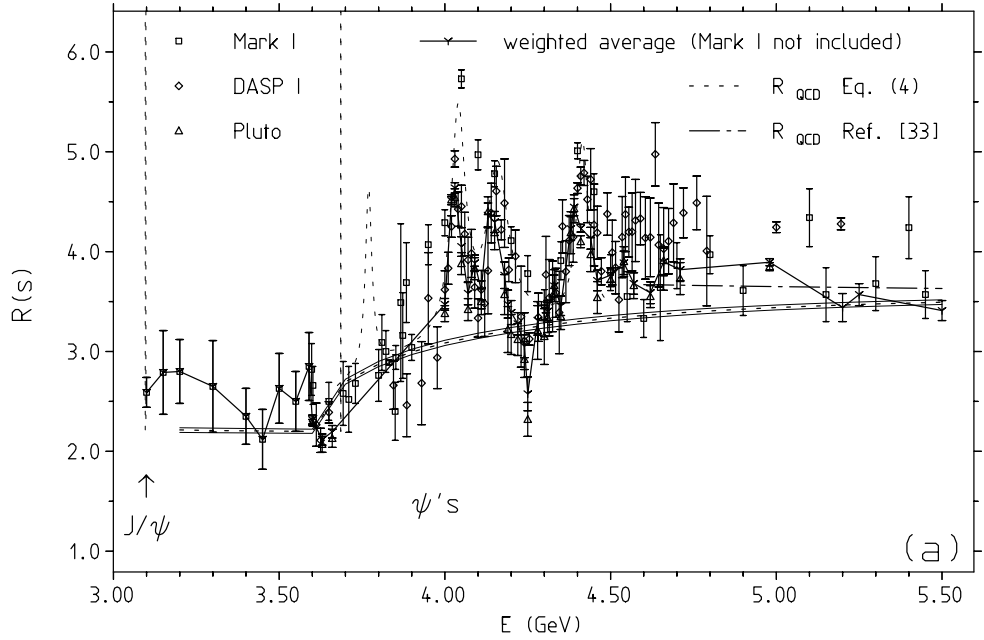
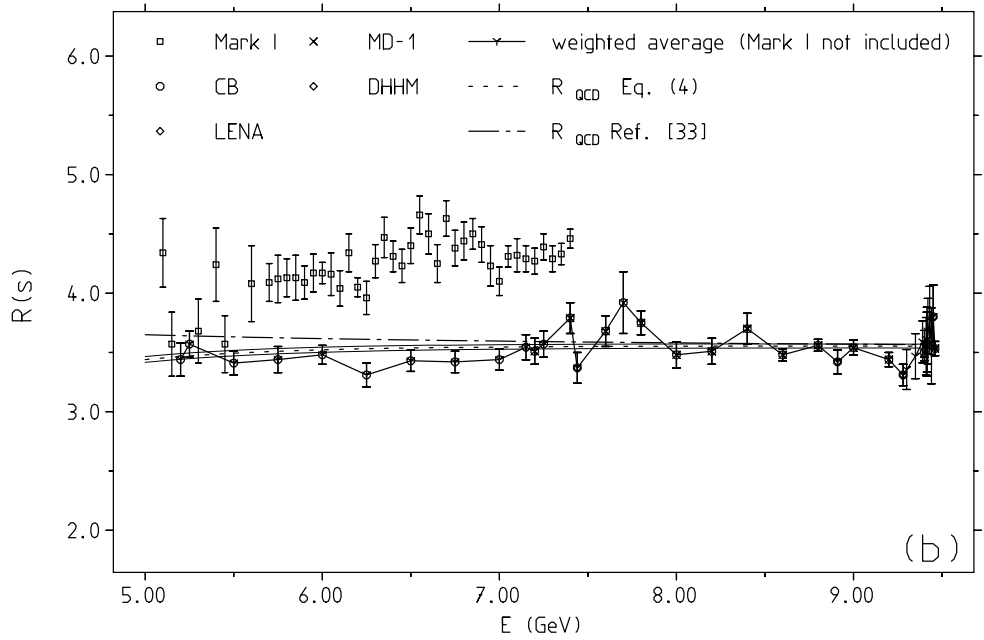
\end{document}